%% file: main.tex
\documentclass[%
reprint,
superscriptaddress,
 amsmath,amssymb,
 aps,
]{revtex4-2}

\usepackage{graphicx}%
\usepackage{dcolumn}%
\usepackage{bm}%
\usepackage{xcolor}
\usepackage{booktabs}
\usepackage{hyperref}
\hypersetup{
	colorlinks=true,
	linkcolor={black},
	citecolor={black},
	urlcolor={black},
}

\begin{document}

\preprint{APS/123-QED}

\title{Deep Learning-based Prediction of Electrical Arrhythmia Circuits from Cardiac Motion: An In-Silico Study}

 \author{Jan Lebert}
 \affiliation{Cardiovascular Research Institute, University of California, San Francisco, USA}

\author{Daniel Deng}
 \affiliation{Cardiovascular Research Institute, University of California, San Francisco, USA}
  \affiliation{Department of Computer Science, University of California, Berkeley, USA}

  \author{Lei Fan}
 \affiliation{College of Engineering, Michigan State University, USA}
 
   \author{Lik Chuan Lee}
 \affiliation{College of Engineering, Michigan State University, USA}

\author{Jan Christoph}
 \homepage{http://cardiacvision.ucsf.edu/}
 \email{jan.christoph@ucsf.edu}
\affiliation{Cardiovascular Research Institute, University of California, San Francisco, USA}
\affiliation{Division of Cardiology, University of California, San Francisco, USA}

\newcommand{\stimes}{{\times}}

\begin{abstract}

\input{sections/abstract}

\begin{description}
\item[Keywords]
Deep Learning, Ventricular Tachycardia, Cardiac Electrophysiology, Ventricular Mechanics
\end{description}

\end{abstract}

\maketitle

\section{Introduction}
\input{sections/introduction}

\section{Methods}

\input{sections/methods}

\section{Results}
\input{sections/results}

\section{Discussion}
\input{sections/discussion}

\section{Conclusions}
\input{sections/conclusions}

\section*{Funding}
This research was funded by the University of California, San Francisco, the National Institutes of Health (DP2HL168071), and the Sandler Program for Breakthrough Biomedical Research, which is partially funded by the Sandler Foundation (to J.C.). This research was also supported through the Academic Hardware Grant program by the NVIDIA Corporation (to J.L. and J.C.).

\section*{Author Contributions}
JL and DD developed the smoothed particle hydrodynamics simulations, the deep learning algorithms and performed the data analysis.
FL and LL performed the FEM simulations.
JL and JC designed the figures.
JC wrote the manuscript.
JL and JC conceived the research.

\section*{Supplemental Data}
Supplementary Videos are available online at: \href{https://cardiacvision.ucsf.edu/videos/}{https://cardiacvision.ucsf.edu/videos/} and at: \href{https://youtube.com/@cardiacvision}{https://www.youtube.com/@cardiacvision}.

\vspace*{10em}

\bibliography{references}

\end{document}

%% file: sections/abstract.tex
The heart's contraction is caused by electrical excitation which propagates through the heart muscle. It was recently shown that the electrical excitation can be computed from the contractile motion of a simulated piece of heart muscle tissue using deep learning.
In cardiac electrophysiology, a primary diagnostic goal is to identify electrical triggers or drivers of heart rhythm disorders. 
However, using electrical mapping techniques, it is currently impossible to map the three-dimensional morphology of the electrical waves throughout the entire heart muscle, especially during arrhythmias such as ventricular tachycardia.
Therefore, the approach to calculate or predict electrical excitation from the hearts motion could be a promising alternative diagnostic approach.
Here, we demonstrate in computer simulations that it is possible to predict three-dimensional electrical wave dynamics from ventricular deformation mechanics using deep learning.
We performed thousands of simulations of focal and reentrant electromechanical activation dynamics in idealized bi-ventricular geometries %
and used the data to train a neural network to analyze the ventricular deformation mechanics and subsequently predict the three-dimensional electrical wave pattern that caused the deformation.
We demonstrate that, next to focal wave patterns, even complicated three-dimensional electrical scroll wave patterns can be reconstructed, even if the network has never seen the particular arrhythmia or heart geometry and was trained on a different electrophysiological or mechanical model.
We show that the deep learning model has the ability to generalize by training it on data generated with the smoothed particle hydrodynamics (SPH) method and subsequently applying it to data generated with the finite element method (FEM).
Predictions can be performed in the presence of scars and with significant heterogeneity. %
Our results suggest that, with adequate training data, deep neural networks could be used to calculate intramural action potential wave patterns from imaging data of the motion of the heart muscle.

%% file: sections/introduction.tex
Ventricular arrhythmias, such as monomorphic or polymorphic ventricular tachycardia (VT) or Torsade de Pointes (TdP), are caused by abnormal three-dimensional electrical wave phenomena which propagate through the thick ventricular heart muscle and initiate irregular, asynchronous contractions.
Cardiac electrophysiologists perform catheter-based mapping of the electrical waves and their interactions with the underlying tissue substrate to determine the origin of heart rhythm disorders.
However, current mapping techniques only provide point-by-point measurements across the heart's surface, while the intramural substrate and electrical wave dynamics remain hidden.
Intramural measurements can be performed with needle catheters, but their use is invasive, provides only localized information, and the highly specialized diagnostics are only performed by a few laboratories. 
Therefore, alternative measurement and visualization methods of intramural electrical activation dynamics are highly needed.

\begin{figure}[htb]
  \centering
  \includegraphics[clip, trim=0.0cm 0.0cm 0.0cm 0.0cm, width=0.45\textwidth]{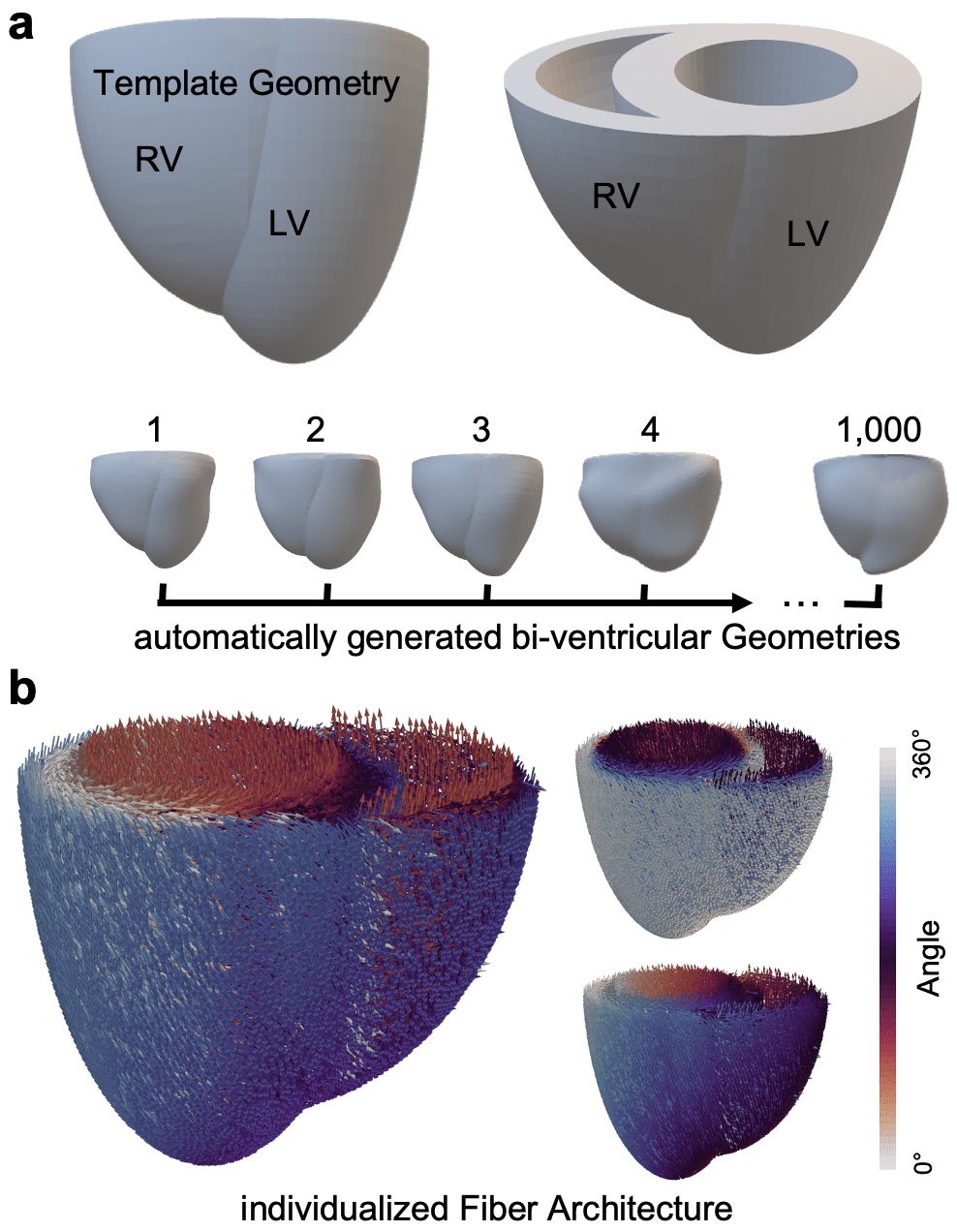}
  \caption{Generation of 1,000 different bi-ventricular heart geometries for generation of electromechanical training dataset consisting of tens of thousands of training samples of ventricular arrhythmias.
  a) Manually designed template geometries (N=10) were automatically deformed using remeshing and free-form deformation algorithms to create 1,000 individual bi-ventricular geometries, see also Supplementary Video 1.
  b) Each bi-ventricular shape comprises its own, unique muscle fiber architecture generated using a rule-based method. Color: Fiber angle in the y-z- (left), x-y- (top right) and x-z-plane (bottom right).}
  \label{fig:geometry_and_fibers}
\end{figure}

In a seminal study by Otani et al. \cite{Otani2010} it was proposed to calculate the electrical wave phenomena from the heart muscle deformation that occurs in response to the electrical excitation. 
More precisely, it was proposed to calculate regions of active stress as an approximation for the electrical waves, because electrical excitation triggers intracellular calcium release, which binds to actin-myosin filaments and promotes their contraction and therefore generates active contractile stress and deformation. %
It was subsequently shown for a planar active stress wave in a simulated cube of myocardial tissue that this inverse problem can be solved using a biophysical model for heart muscle tissue and inverting this model to solve for active stress.
These findings were early indications that the deformation of the heart wall could reveal electrical wave patterns.
However, cardiac arrhythmia circuits often involve more complex electrical wave mophologies, such as reentrant waves.
Subsequently, Lebert et al. \cite{Lebert2019} and Beam et al. \cite{Beam2020, Beam2020a} showed that also more complicated rotating electrical wave patterns, such as spiral waves and even three-dimensional scroll waves, can be reconstructed from deformation in simulated slabs of myocardial tissue using data assimilation.
With both techniques, the electrical waves were reconstructed directly, skipping active stress, by making generalized assumptions about the coupling between electrics and mechanics.
However, these assumptions involved biophysical formulations for either of the electrical wave dynamics, coupling between electrics and mechanics, the tissue's contractility and elastic behavior, and information such as the muscle fiber organization, which is not known per se when imaging the heart. 

More recently, we showed that a purely data-driven deep learning-based approach can be used to solve the inverse mechano-electrical problem, even with very complicated three-dimensional electrical scroll wave dynamics, noise and low spatial resolutions \cite{Christoph2020}, the approach outperforming the previous physics-based approaches \cite{Lebert2019, Beam2020, Beam2020a}.
The deep learning-based predictions of the electrical wave dynamics were visually indistinguishable from the ground-truth, even though the scroll wave patterns were fine-scaled, rapidly changing and chaotic, and the tissue retained an orthotropic muscle fiber organization which was not known to the algorithm.
Next to its computational efficiency, the data-driven approach has also the advantage that no biophysical laws or model equations need to be known as in \cite{Otani2010,Lebert2019,Beam2020}.
Instead, a convolutional encoding-decoding neural network was trained with pairs of electrical and mechanical data, which subsequently learned to predict electrical wave patterns directly from mechanics, %
skipping the biophysics of the excitation-contraction coupling machinery entirely.

\begin{figure}[b]
  \centering
  \includegraphics[clip, trim=0.0cm 0.0cm 0.0cm 0.0cm, width=0.42\textwidth]{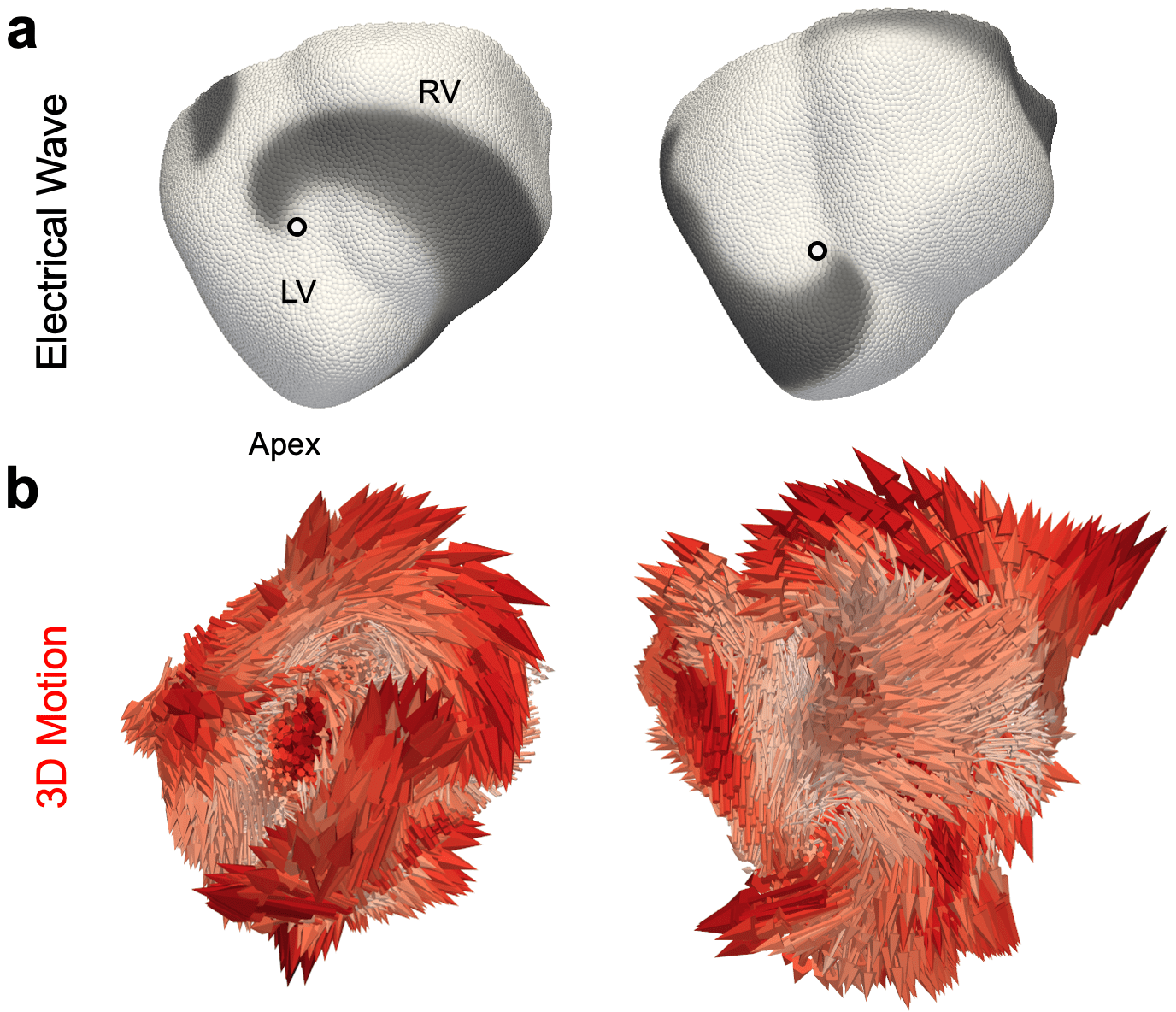}
  \caption{Computer simulation of ventricular deformation during reentrant arrhythmia. 
  a) Electrical wave (dark gray: depolarized tissue, light gray: resting tissue) rotating around a core (white dot) or 'reentering' the ventricles.
  b) Corresponding ventricular deformation caused by the electrical wave. Motion displayed as three-dimensional vectors (red), indicating instantaneous motion with respect to a previous deformed state.
  In this study, we ask the question: does the mechanical motion that was caused by the electrical excitation contain sufficient information to be able to reconstruct the electrical wave pattern?}
  \label{fig:electromechanical_rotor}
\end{figure}

Based on these preliminary studies, in this study, we aimed to investigate whether deep learning could also be applied to solve the inverse mechano-electrical problem in more realistic and arbitrarily-shaped tissue geometries, such as different ventricular geometries that could stem from different patients, and in different tissues with significantly different underlying electrophysiology and mechanics. %
We therefore performed thousands of electromechanical computer simulations with bi-ventricular heart geometries, see Figs.~\ref{fig:geometry_and_fibers} and \ref{fig:electromechanical_rotor}, with and without scars.
To obtain a rich and diverse training dataset of ventricular arrhythmias, we randomly chose different electrophysiological and mechanical models and varied simulation parameters influencing the wavelength and other properties of the electrical waves, for instance, as well as the tissue's contractile and elastic behavior.
We then trained a neural network with the data and explored whether it can learn the relationship between three-dimensional electrical waves and ventricular deformation, and subsequently devise a unique, generalizable and robust mapping between electrics and mechanics, despite the complexity of the problem.

\begin{figure*}[ht]
  \centering
  \includegraphics[clip, trim=0.0cm 0.0cm 0.0cm 0.0cm, width=0.98\textwidth]{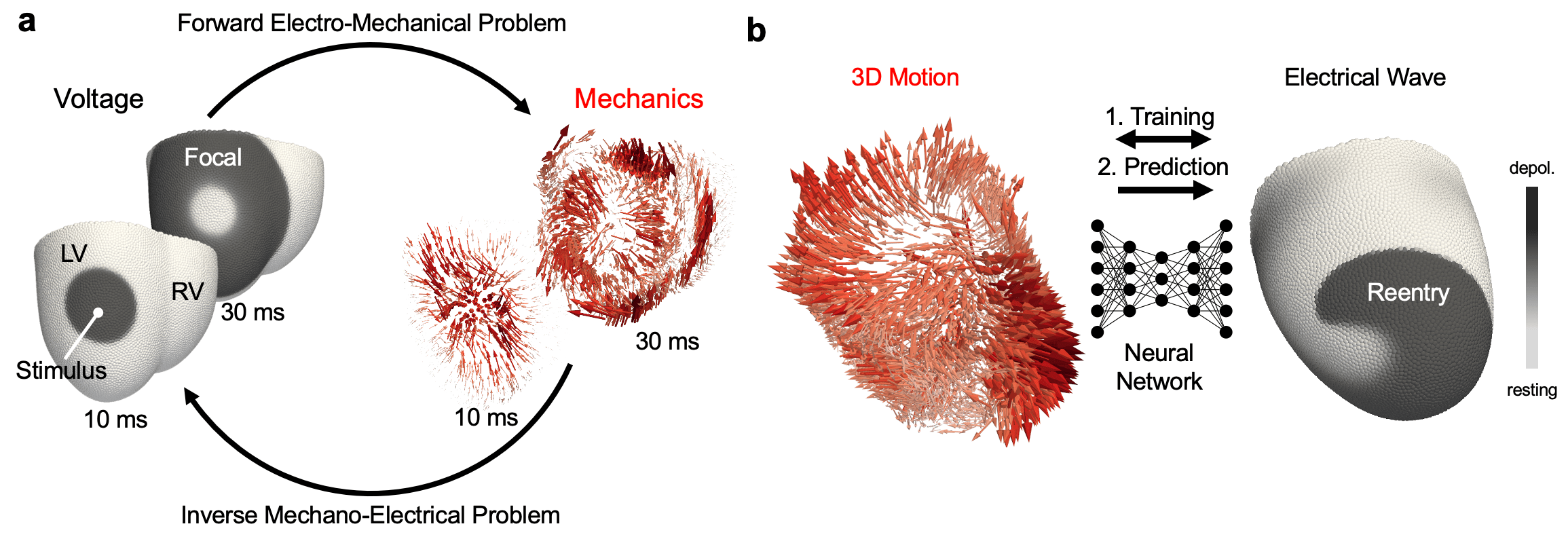}
  \caption{Deep learning-based prediction of electrical arrhythmia circuits from mechanical deformation in simulations of the heart's ventricles. 
  The possibility to compute action potential waves from cardiac tissue motion could help improve diagnosing heart rhythm disorders and other heart disease.
  a) Forward and inverse problem: In the forward electro-mechanical problem, computer simulations are used to compute electrical waves which propagate through the heart muscle and initiate contractile motion. The example shows a focal wave that originates from a point, propagates outwards and induces focal-like contractions (tissue motion is indicated by red three-dimensional displacement vectors). 
  In the inverse mechano-electrical problem, we aim to compute electrical waves from tissue motion.
  b) A neural network is used to solve the inverse mechano-electrical problem, see also \cite{Christoph2020}. 
  After training with many pairs of mechanical input and electrical target data it can perform predictions with previously unseen data.}
  \label{fig:inverse-problem}
\end{figure*}

Experimental work has provided evidence for a strong correlation between electrical and mechanical spatio-temporal patterns in the heart \cite{Wyman1999, Provost2011, Christoph2018, Maffessanti2020}. 
For example, it was shown during ventricular arrhythmias that focal or rotational electrical wave phenomena visible on the heart surface cause focal or rotational mechanical deformation patterns within the heart wall \cite{Christoph2018, Molavi2022}. 
These observations motivate our work and suggest that 
our proposed approach could be applied to process such mechanical patterns in practice.

%% file: sections/methods.tex
We developed and trained a sparse encoding-decoding convolutional neural network using electromechanical computer simulations to predict electrical excitation waves from ventricular tissue deformation.

\subsection{Computer Simulations of Bi-Ventricular Electro-Mechanics}
\begin{figure*}[htb]
  \centering
  \includegraphics[clip, trim=0.0cm 0.0cm 0.0cm 0.0cm, width=0.96\textwidth]{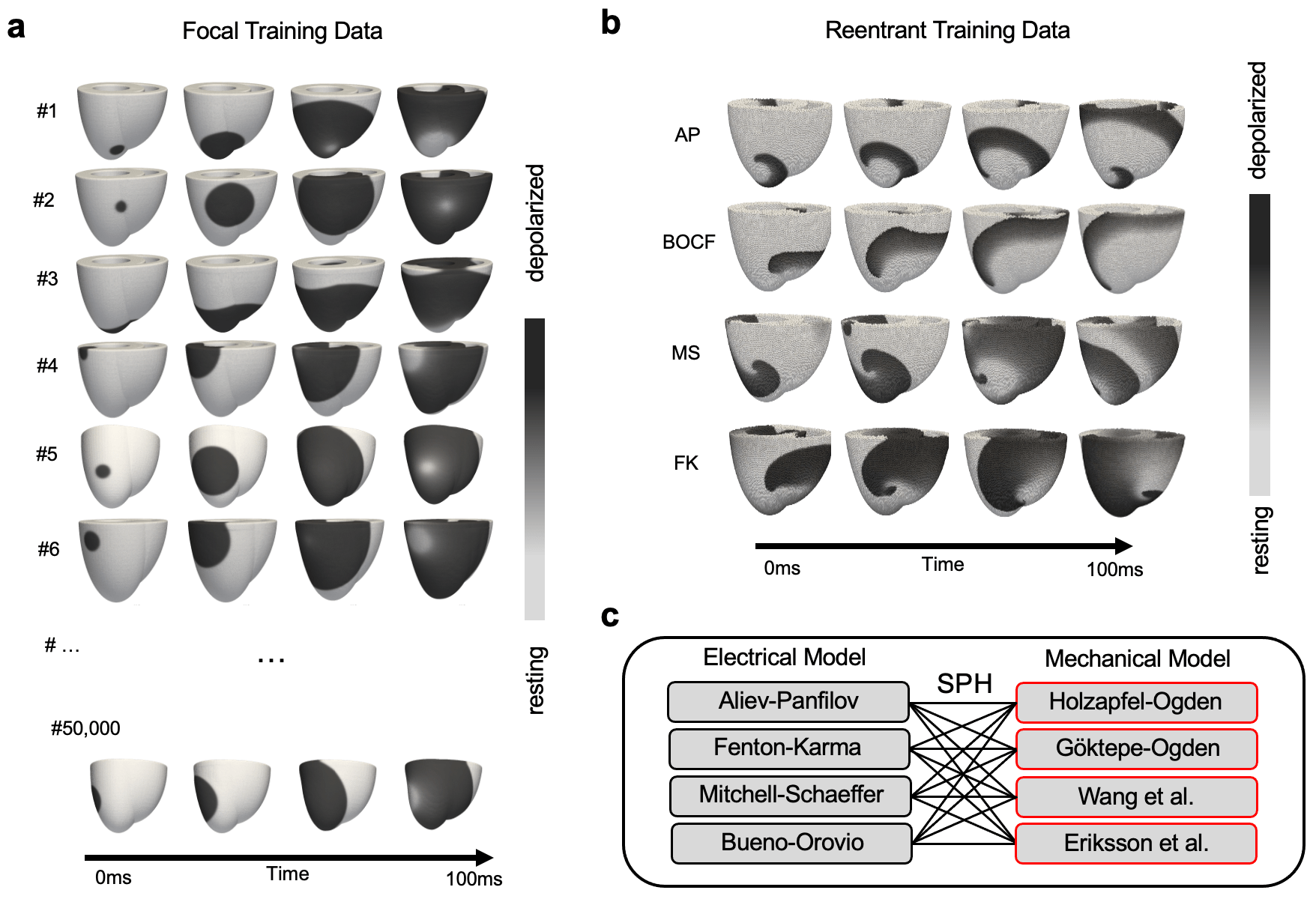}
  \caption{Simulated training data of focal and reentrant ventricular arrhythmias generated with electromechanical computer simulations. 
  a) Focal training data was generated using a single S1 stimulation pulse in a random location in the ventricles.
  Focal waves originate from a single point source. 
  We performed up to $1,000$ different simulations with a single randomly placed stimulation pulse, randomly selected electrophysiological and mechanical models, see c), and randomly chosen parameters varying properties of the electrical waves, tissue stiffness and contractile strength, among others.
  b) Reentrant training data was generated using a S1-S2 stimulation protocol in which a S1 stimulation pulse was first applied in a random location and then a S2 stimulation pulse was applied in the vulnerable phase of the wave causing wavebreak and reentry.
  Reentrant waves rotate around a core or an anatomical site (such as a scar or an occlusion).
  Different reentrant scroll wave dynamics with the Aliev-Panfilov \cite{AlievPanfilov1996} (AP), Bueno-Orovio-Cherry-Fenton \cite{BuenoOrovioCherryFenton2008} (BOCF), Mitchell-Schaeffer \cite{MitchellSchaeffer2003} (MS) and Fenton-Karma \cite{FentonKarma1998} (FK) models.
  c) Combinations of electrophysiological and mechanical models used for the simulations in a,b).}
  \label{fig:training_data}
\end{figure*}

\begin{figure*}[htb]
  \centering
  \includegraphics[clip, trim=0.0cm 0.0cm 0.0cm 0.0cm, width=0.98\textwidth]{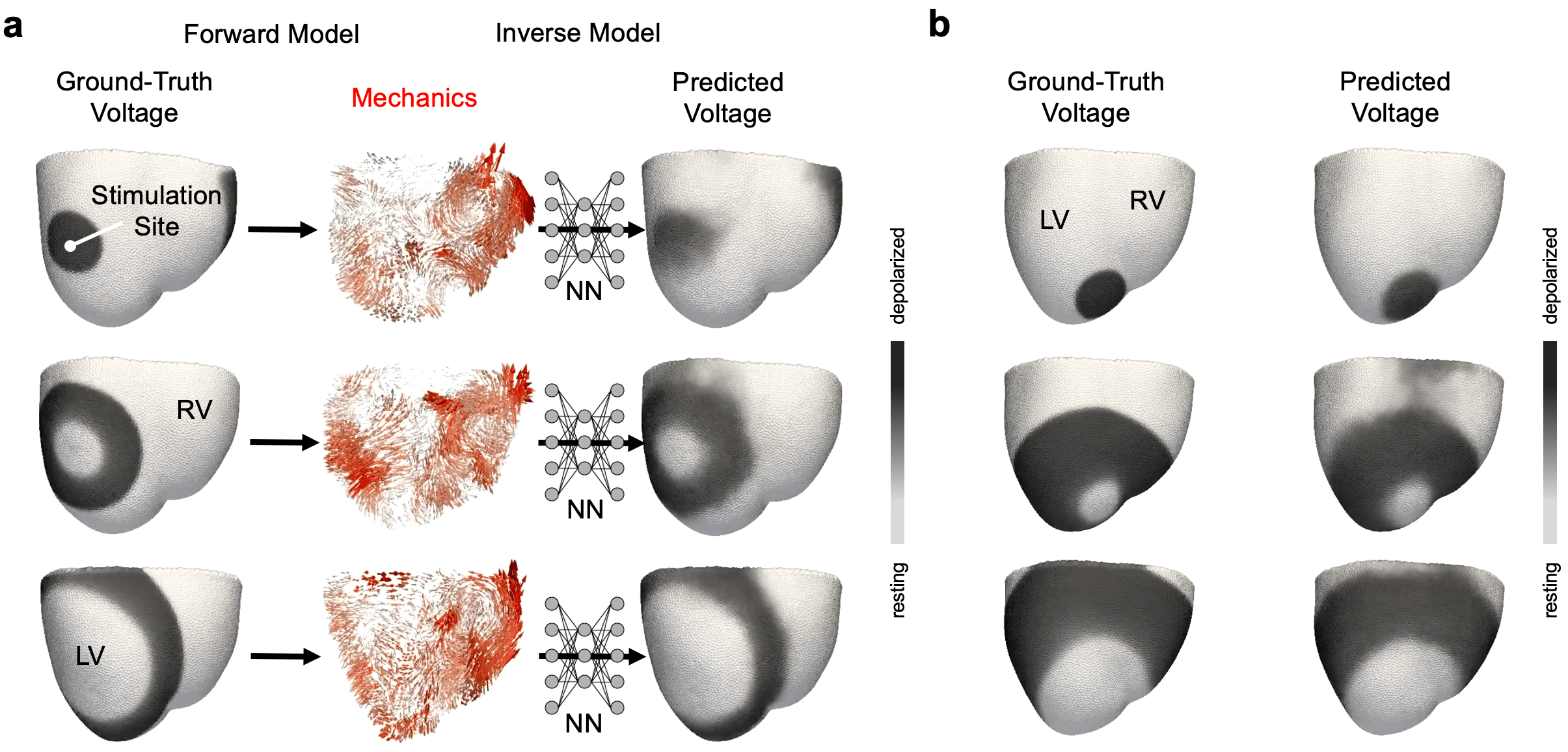}
  \caption{Deep learning-based predictions of focal electrical (action potential) waves from ventricular mechanical deformation.
  a) Representative example of a focal voltage wave (left, ground-truth) originating from the left ventricle causing contractions and subsequent deformation of both ventricles (center, red displacement vectors), which is analyzed by a neural network (NN) to predict the electrical wave pattern (right). 
  Each prediction was performed individually and the sequence of predictions yields a temporally consistent focal wave pattern, see Supplementary Videos 2 and 3.
  b) Second example with ground-truth and predicted electrical focal wave originating from apex region.
  The data was not seen by the neural network during training (neither the particular bi-ventricular geometry nor the focal wave).}
  \label{fig:predictions_focal}
\end{figure*}

We performed computer simulations of electromechanical wave dynamics in idealized bi-ventricular geometries with muscle fiber anisotropy, see Fig.~\ref{fig:geometry_and_fibers}.
The simulations reproduced two fundamental types of electromechanical activation dynamics in the ventricles which we assume to underlie ventricular tachyarrhythmias: either i) focal or ii) reentrant wave dynamics, see also Figs.~\ref{fig:training_data}a,b), \ref{fig:predictions_focal} and \ref{fig:predictions_reentrant}.
Both focal or reentrant wave patterns can underlie either monomorphic VT, polymorphic VT or Torsade de Pointes (TdP). 
Monomorphic VT (mVT) was simulated as being caused by either a single focal wave originating from a point source located in a random position in the ventricles or a single stable macro-reentrant wave or scroll wave, respectively.
Polymorphic VT (pVT) was simulated as a single meandering scroll wave or multiple scroll waves, respectively.
The latter dynamical state is similar to TdP or ventricular fibrillation (VF).
Reentrant waves were induced using an S1-S2 stimulation protocol with a first S1 stimulus in a random location and one or two additional S2 stimuli in random locations timed to occur in the vulnerable window to cause wavebreak. 
This protocol was repeated to induce multiple scroll waves.

Throughout this study, we refer to electrical models and mechanical models.
The electrical models are phenomenological numerical models for action potential waves in cardiac tissue.
For simplicity, we refer to mechanical models, e.g. in sections \ref{sec:methods:SPH}, Fig.~\ref{fig:training_data} and sections \ref{sec:results_focal}-\ref{sec:results_reentrant}, when we actually refer to different mechanical material properties.
We employ only forward electromechanical coupling using phenomenological laws, coupling the mechanics to the electrics, omitting feedback effects of the mechanics onto the electrics, see sections \ref{sec:methods:SPH} and \ref{sec:methods:FEM}, respectively.

\subsubsection{Smoothed Particle Hydrodynamics Simulations}
\label{sec:methods:SPH}
To be able to generate large quantities of training data with tens of thousands of training samples, we used an efficient numerical electromechanical model based on the smoothed particle hydrodynamics (SPH) method implemented in the general-purpose open-source SPHinXsys framework developed by Zhang et al. \cite{Zhang2021, Zhang2021b}, which also provides a cardiac-specific simulation framework.
The SPH method is a meshfree Lagrangian numerical method that discretizes the simulation domain using particles and offers the ability to handle complex tissue geometries. 
We modified and extended this simulation framework to be able to address the research questions in this study.
We performed $2,000$ simulations in total: $1,000$ simulations with $1,000$ different unique bi-ventricular geometries with unique muscle fiber anisotropy without scars, and another $1,000$ simulations with $1,000$ different bi-ventricular geometries with scars, respectively.
The geometries and corresponding fiber architectures were created automatically, as described in section \ref{sec:methods:ventricular_geometries} and illustrated in Fig.~\ref{fig:geometry_and_fibers}, 

Electrical action potential wave patterns were modelled using 4 different phenomenological electrophysiological models (Aliev-Panfilov \cite{AlievPanfilov1996}, Fenton-Karma \cite{FentonKarma1998}, Mitchell-Schaeffer \cite{MitchellSchaeffer2003}, Bueno-Orovio-Cherry-Fenton \cite{BuenoOrovioCherryFenton2008}). 
We selected one of the models randomly during the initialization of each simulation. 
The electrical models describe reaction-diffusion dynamics using coupled partial differential equations of the form:
\begin{eqnarray}
    \partial_t u & = & \nabla \cdot(\mathbf{D} \nabla u) +  f(u,v) \\
    \partial_t v & = & g(u,v) 
\end{eqnarray}
where $u$ is the excitatory variable, $v$ is the slow/gating variable (some models have several slow/gating variables $v,w$ etc.), $\nabla \cdot(\mathbf{D} \nabla u)$ is a term for the diffusion, $\mathbf{D}$ is the anisotropic diffusion tensor, and $f(u,v)$, $g(u,v)$ are the local excitatory kinetics.
The parameter values for each model are shown in Table \ref{tab:electrics_parameters}.
Each model produces distinctly different wave patterns, see Fig.~\ref{fig:training_data}b).
However, qualitatively, all models produce electrical wave patterns typically observed in cardiac muscle tissue: focal and reentrant scroll waves or compositions thereof, see Fig.~\ref{fig:training_data}a,b).
Fig.~\ref{fig:training_data}b) illustrates the quantitative differences between the electrophysiological models for scroll wave patterns. 
Note that the scroll wave dynamics are a snapshot of the model dynamics for a given parameter set and that each model can exhibit different and more diverse dynamics.
For focal stimulations, the interval between stimuli was chosen at random from the range 0.5 s to 1.3 s (simulation time).
In the electrical part of the simulations, we varied the isotropic and anisotropic diffusion coefficients, scaled the diffusion tensor, the slow/gating variables and introduced spatial heterogeneity, see sections \ref{sec:methods:heterogeneity}-\ref{sec:methods:scars}.

\begin{figure*}[htb]
  \centering
  \includegraphics[clip, trim=0.0cm 0.0cm 0.0cm 0.0cm, width=0.95\textwidth]{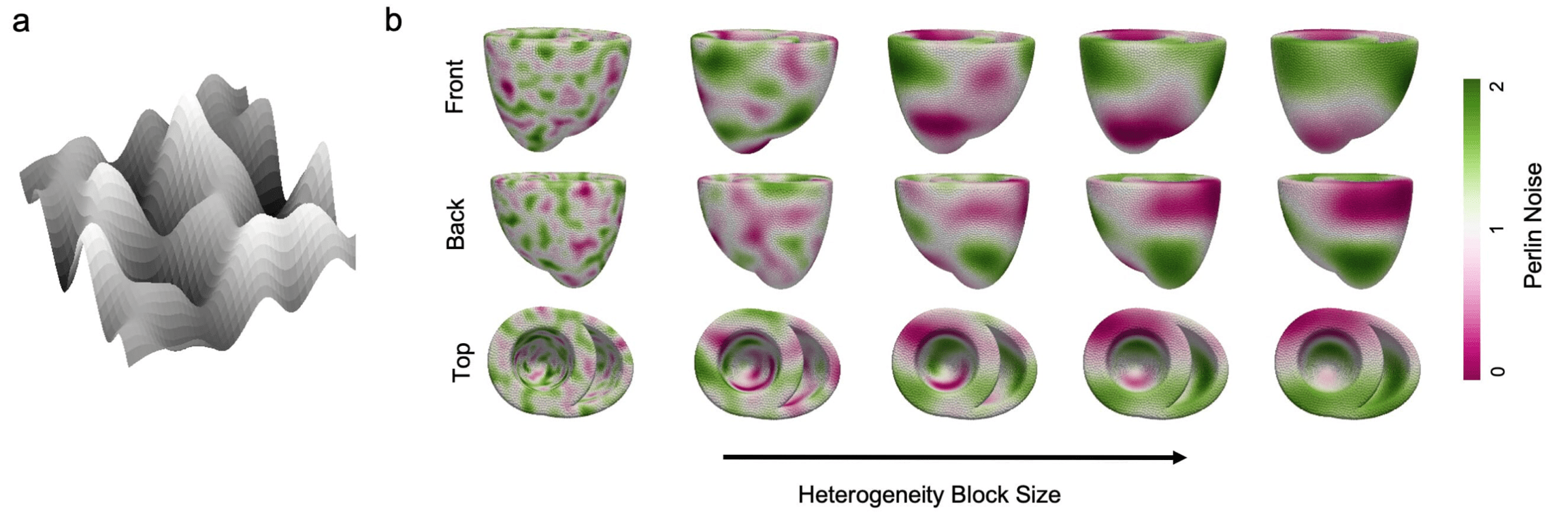}
  \caption{Automatic generation of spatial heterogeneity using Perlin noise ($z$-axis: intensity). 
  a) Example of Perlin noise in two spatial dimensions.
  b) Spatial heterogeneity randomly initiated in each simulation.
  We modulated the block size or spatial frequency as well as the intensity of the Perlin noise.
  The random noise modulated electrical and mechanical parameters, which caused more and less noise to the dynamics.
  The particular realization of the noise or noise pattern in each simulation was different from the noise pattern used for scar generation, see Fig.~\ref{fig:scars}.
  }
  \label{fig:heterogeneity}
\end{figure*}

\begin{table}[htb]
\centering
\resizebox{0.47\textwidth}{!}{%
\begin{tabular}{l l l l l l l l l l l l} \toprule
   \multicolumn{12}{c}{Aliev-Panfilov (AP) model} \\ \midrule
   $a$ & 0.2 & $b$ & 0.2 & $k$ & 8 & $\mu_1$ & 0.2 & $\mu_2$ & 0.3 & $\epsilon$ & 0.002\\ \midrule
   \multicolumn{12}{c}{Fenton-Karma (FK) model} \\ \midrule
   $\tau_v^+$ & 10 & $\tau_{v1}^-$ & 18.2 & $\tau_{v2}^-$ & 18.2 & $\tau_w^+$ & 1020 & $\tau_{w}^-$ & 80 & $\tau_d$ & 1/5.8\\
   $\tau_0$ & 12.5 & $\tau_r$ & 130 & $\tau_{\text{si}}$ & 127 & $k$ & 10 & $u_c$ & 0.13 & $u_c^{\text{si}}$ & 0.85 \\
   $u_v$ & --\\\midrule
   \multicolumn{12}{c}{Mitchell-Schaeffer (MS) model} \\ \midrule
   $\tau_{\text{in}}$ & 0.3\,ms & $\tau_{\text{out}}$ & 6\,ms & $\tau_{\text{open}}$ & 120\,ms & $\tau_{\text{close}}$ & 150\,ms & $u_{\text{gate}}$ & 0.13 & &\\ \midrule
   \multicolumn{12}{c}{Bueno-Orovio-Cherry-Fenton (BOCF) model} \\ \midrule
   $u_0$ & 0 & $u_u$ & 1.58 & $\theta_v$ & 0.3 & $\theta_w$ & 0.015 & $\theta_v^-$ & 0.015 &
   $\theta_o$ & 0.006 \\
   $\tau_{v1}^-$ & 60 & $\tau_{v2}^-$ & 1150 & $\tau_{v}^+$ & 1.4506 & $\tau_{w1}^-$ & 70 & $\tau_{w2}^-$ & 20 & $u_w^-$ & 0.03 \\
   $\tau_w^+$ & 280 & $\tau_{\text{fi}}$ & 0.11 & $\tau_{o1}$& 6 & $\tau_{o2}$ & 6 & $\tau_{so1}$ & 43 & $\tau_{so2}$ & 0.2\\
   $k_{so}$ & 2 & $u_{so}$ & 0.65 & $\tau_{s1}$ & 2.7342 & $\tau_{s2}$ & 3 & $k_s$ & 2.0994 & $u_{s}$ & 0.9087\\ $\tau_{si}$ & 2.8723 & $\tau_{w\infty}$ & 0.07 & $w_{\infty}^{\star}$ & 0.94\\
   \bottomrule
\end{tabular}
}
\caption{Parameter values for the 4 electrophysiological models (AP \cite{AlievPanfilov1996}, FK \cite{FentonKarma1998}, MS \cite{MitchellSchaeffer2003}, BOCF \cite{BuenoOrovioCherryFenton2008}) used for training data generation.}
\label{tab:electrics_parameters}
\end{table}

The ventricular mechanics were modeled following the approach by \textcite{Nash2004} by decomposing the Piola-Kirchhoff stress tensor $\textbf{P}$ as the sum of passive ($\textbf{P}_p$) and active ($\textbf{P}_{a}$) stress components. The passive stress is derived using a strain-energy function $W$, which accounts for the anisotropic and nonlinear mechanical properties of the myocardium:
\begin{align}
    W = U + W_{\text{iso}} + W_{\text{aniso}}\text{,}
\end{align}
where $U$ represents a volumetric function, and $W_{\text{iso}}$ and $W_{\text{aniso}}$ denote the isotropic and anisotropic volume-preserving functions, respectively.
The volumetric function $U$ was chosen as $U(J) = \mu_k \ln(J)^2/2$, where $J=\text{det} \mathbf{F}$ is the volume ratio defined as the determinant of the deformation gradient tensor $\mathbf{F}$, and $\mu_k$ is the bulk modulus that acts as a penalty parameter to enforce incompressibility.
The isotropic and anisotropic strain energy functions, $W_{\text{iso}}$ and $W_{\text{aniso}}$, are defined using the constitutive law proposed by \textcite{Holzapfel2009}:
\begin{align}
    W_{\text{iso}} &= \frac{a}{2b} \text{e}^{b(I_1-3)}\text{,}\\
    W_{\text{aniso}} &= \sum_{i=\text{f},\text{s}}\frac{a_i}{2b_i}\left(\text{e}^{b_i(I_{4i}-1)^2}-1\right) + \frac{a_\text{fs}}{2b_\text{fs}}\left(\text{e}^{b_\text{fs}I_{8\text{fs}}}-1\right)\text{.}
\end{align}
Here, $a$ and $b$ are parameters related to the matrix response, $a_\text{f}$ and $b_\text{f}$ account for the myocardial fiber contributions, $a_\text{s}$ and $b_\text{s}$ denote the parameters for the fiber sheet contributions, and $a_\text{fs}$ and $b_\text{fs}$ characterize the shear effects in the sheet plane. For definitions of the invariants $I_1, I_{4\text{f}}, I_{4\text{s}}$, and $I_{8\text{fs}}$, as well as the formulation of the passive stress tensor $\textbf{P}_p$ see \textcite{Zhang2021b}.
The active stress component modeling the internal active contraction stress was defined as $\textbf{P}_a = T_a(u) \textbf{F} \textbf{f}_0\otimes\textbf{f}_0$, where $T_a(u)$ represents the active fiber tension in dependence of the excitatory variable $u$ from eq. (1) and $\textbf{f}_0$ is the local fiber direction. 
The evolution of the active fiber tension or active stress is described as \cite{Nash2004}:
\begin{align}
    \partial_t T_a &= \epsilon_{T}(u) (k_{T} u  - T_a)\text{,}\label{eq:ta}
\end{align}
where $\epsilon_{T}(u)$ is a delay function controlling the rate of activation and relaxation. 
We employed the smoothly varying form of $\epsilon_{T}(u)$ proposed by \textcite{Goktepe2010}:
\begin{align}
  \epsilon_{T}(u) &= \epsilon_0 + (\epsilon_\infty - \epsilon_0) \exp{\left(-\exp{\left(-\xi_T (u - u_0)\right)}\right)}\text{,}\label{eq:epsilont:goektepe}
\end{align}
where $\epsilon_0$ and $\epsilon_\infty$ represent the lower and upper bounds of $\epsilon_{T}(u)$, respectively, while $\xi_T$ controls the rate of transition between these values around $u_0$. We adopt the corrected relation $\epsilon_\infty < \epsilon_0$, as noted by \textcite{Eriksson2013}, and chose $\epsilon_\infty = 0.1$, $\epsilon_0=1.0$, $\xi_T=100$, and $u_0 = 0.05$.
To generate a diverse dataset, we varied the following parameters for each simulation: (1) the bulk modulus $\mu_k$, (2) the material parameters of the Holzapfel-Ogden constitutive law ($a$, $b$, $a_\text{f}$, $b_\text{f}$, $a_\text{s}$, $b_\text{s}$, $a_\text{fs}$, and $b_\text{fs}$), and (3) the maximum active force $k_T$.
First, the bulk modulus $\mu_k$ was set as a function of the Young's modulus $E$ using $\mu_k = E/(3(1-2\nu))$ with the Poisson's ratio $\nu = 0.4995$. We choose $E$ from a uniform distribution between $5$\,kPa to $15$\,kPa for each simulation.
Second, the passive material parameters $a$, $b$, $a_\text{f}$, $b_\text{f}$, $a_\text{s}$, $b_\text{s}$, $a_\text{fs}$, and $b_\text{fs}$ were chosen in two ways: a base value parameter set from four different sources, and a scaling factor $k_a$ which scales $a$, $a_\text{f}$, $a_\text{s}$, $a_\text{fs}$. Table \ref{tab:mechanics_parameters} lists the base value parameter sets from \textcite{Holzapfel2009, Goktepe2010b, Wang2012, Eriksson2013}, which stem from different fits on shear test data of porcine ventricular myocardium \cite{Dokos2002}.
For each simulation, we randomly selected one parameter set and choose the scaling factor $k_a$ randomly from the range $0.8$ to $1.2$.

\begin{table}[htb]
   \centering
   \resizebox{0.47\textwidth}{!}{%
   \begin{tabular}{@{}l *{8}{l} @{}} \toprule
  Source  & \multicolumn{1}{c}{$a$~[kPa]} & \multicolumn{1}{c}{$b$}    & \multicolumn{1}{c}{$a_\text{f}$~[kPa]} & \multicolumn{1}{c}{$b_\text{f}$} & \multicolumn{1}{c}{$a_\text{s}$~[kPa]} & \multicolumn{1}{c}{$b_\text{s}$} & \multicolumn{1}{c}{$a_\text{fs}$~[kPa]} & \multicolumn{1}{c}{$b_\text{fs}$} \\ \midrule
   Holzapfel \cite{Holzapfel2009}   & 0.059     & 8.023  & 18.472             & 16.026       & 2.481              & 11.120       & 0.216               & 11.436        \\
   Goktepe \cite{Goktepe2010b}    & 0.496     & 7.209  & 15.196             & 20.417       & 3.283              & 11.176       & 0.662               & 9.466         \\
   Wang \cite{Wang2012}        & 0.2362    & 10.810 & 20.037             & 14.154       & 3.7245             & 5.1645       & 0.4108              & 11.300        \\
   Eriksson \cite{Eriksson2013}    & 0.333     & 9.242  & 18.535             & 15.972       & 2.564              & 10.446       & 0.417               & 11.602        \\ \bottomrule
   \end{tabular}
   }
   \caption{Different material parameter choices for the Holzapfel-Ogden constitutive model \cite{Holzapfel2009}, derived from different shear tests with porcine ventricular myocardium \cite{Dokos2002}.}
   \label{tab:mechanics_parameters}
\end{table}

\begin{table}[htb]
  \centering
  \resizebox{0.46\textwidth}{!}{
  \begin{tabular}{l l l l l l} \toprule
      \multicolumn{6}{c}{Geometry Generation} \\ \midrule
      Base Geometry & $N_{\text{deform}}$ & $\sigma_{\text{deform}}$ & $\theta_{\text{epi}}$ & $\theta_{\text{endo}} - \theta_{\text{epi}}$ & \\
      $\{\#1,\ldots, \#13\}$ & 8 to 20 & 0.05 to 0.3 & $-90^{\circ}$ to $-60^{\circ}$ & $80^{\circ}$ to $120^{\circ}$ & \\ \midrule 
      \multicolumn{6}{c}{Electrophysiology} \\ \midrule
      Model & Stimulation & $D_{\text{iso}}$ & $D_{\text{ratio}}$ & $\xi_h$ & $p_{\text{scar}}$ \\
      \{AP, FK, MS, BOCF\} & \{S1, S1-S2\} & 0.2 to 0.4 (S1) & 2.0 to 4.0 & 0 to 0.3 & 0\,\% to 20\,\% \\
      & & 0.1 (S1-S2) & & & \\ \midrule
      \multicolumn{6}{c}{Mechanics} \\ \midrule
      Passive Parameters & $k_a$ & $k_{T}$ & $E$ \\
      $\{\#1, \ldots, \#4\}$ & 0.8 to 1.2 & 80 to 120\,kPa & 5 to 15\,kPa \\ \midrule
      \multicolumn{6}{c}{System} \\ \midrule
      $\Delta t_{\text{save}}$ \\
      \multicolumn{3}{l}{13\,ms to 32\,ms (corresponds to 31\,Hz to 77\,Hz)} \\
      \bottomrule
  \end{tabular}
  }
  \caption{Summary of the parameter ranges used to generate the simulation dataset. 
  For all numerical ranges we used a uniform distribution.
  In total, 10-15 parameters were varied depending on the type of simulation.}
  \label{tab:parameters}
\end{table}

\subsubsection{Finite Element Method Simulations}
\label{sec:methods:FEM}
Additional electromechanical simulations (N=6) were carried out using the finite element method (FEM) \cite{Fan2022,Arumugam2019} implemented in the open-source library FEniCS \cite{Logg2012}. 
Simulations were carried out in bi-ventricular geometries which were distinct from the bi-ventricular geometries used with the SPH method described in the previous section \ref{sec:methods:SPH}.
A different simulation-specific implementation for the creation of the ventricular muscle fiber architecture was used than in the SPH simulations, but the fiber architecture was also helical orthotropic with a total angle of rotation of $\theta = 120^{\circ}$ from endo- to epicardium, see also section \ref{sec:methods:ventricular_geometries}.
The FEM methodology was not known to the team members developing the deep learning methodology, the sourcecode was not made available, and a brief description of the modelling details was purposefully only made available upon writing of the manuscript.
Focal electrical waves were simulated using the Aliev-Panfilov model \cite{AlievPanfilov1996} based on the mono-domain formulation as described in \cite{Fan2022,Arumugam2019}. 
Specifically, electrical propagation of the focal waves was controlled by an anisotropic electrical conductivity tensor. 
The FEM simulations were performed at two different spatial resolutions (low, very low) and in general at much lower resolutions than the SPH simulations, see also Fig.~\ref{fig:fem}a).

\subsubsection{Automatic Generation of Unique Ventricular Geometries with Muscle Fiber Architecture}
\label{sec:methods:ventricular_geometries}
Unique idealized bi-ventricular geometries were created automatically for each individual simulation by selecting $1$ out of $10$ different template geometries and randomly deforming this geometry into a unique new geometry, see Fig.~\ref{fig:geometry_and_fibers}a).
The $10$ template bi-ventricular geometries were created manually with computer-aided design software.
Each template had different wall thicknesses, diameter, height, offsets between left and right ventricles, etc.
First, the template geometry mesh was remeshed to achieve a uniform distribution of vertices using a Voronoi clustering technique \cite{Valette2008} implemented in the pyacvd library \cite{Kaszynski2022}. 
The geometry was then deformed using a free-form deformation (FFD) algorithm \cite{Sederberg1986} to smoothly transform the remeshed template geometry into a new, distinctive shape using a random deformation field, see Fig.~\ref{fig:geometry_and_fibers}a) and also Supplementary Video 1.
FFD utilizes a lattice-based control structure to manipulate the shape of the underlying object in a smooth and continuous manner by operating on the vertices of the mesh. 
The deformation process is parameterized by a lattice control structure of dimensions $N_{\text{deform}} \times N_{\text{deform}} \times N_{\text{deform}}$, where $N_{\text{deform}}$ represents the number of control points in each direction. 
We created a random deformation by modifying the lattice control point positions, perturbing them in all directions using Gaussian noise with mean $0$ and standard deviation $\sigma_{\text{deform}}$. 
The values of $N_{\text{deform}}$ and $\sigma_{\text{deform}}$ are both randomly chosen, see Table \ref{tab:parameters}.
The FFD was performed using the PyGeM library \cite{Tezzele2021}, and the resulting geometry was refined using the MeshFix algorithm \cite{Attene2010} to ensure a mesh without self-intersections.
Supplementary Video 1 shows this process for one template geometry.

\begin{figure}[htb]
  \centering
  \includegraphics[clip, trim=0.0cm 0.0cm 0.0cm 0.0cm, width=0.49\textwidth]{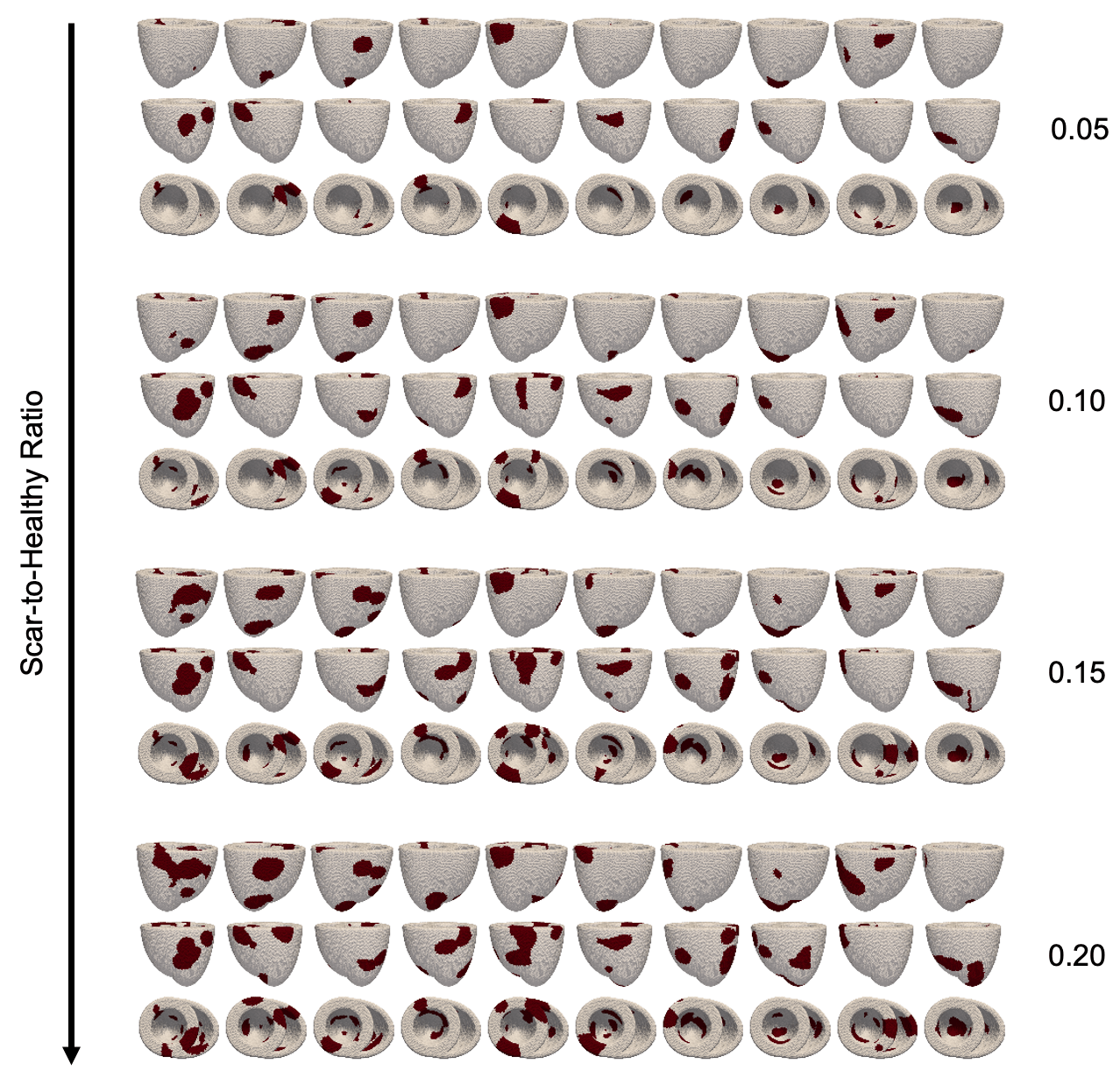}
  \caption{Scar tissue generated for each simulation. 
  The non-conducting and non-contracting tissue was placed in random locations in the ventricles. 
  The block size was varied and the scar-to-healthy tissue ratio or scar density $p_{scar}$ was varied in each simulation from $0$ to $0.2$.
 The particular realization of the Perlin noise in each simulation was different from the Perlin noise used for spatial heterogeneity generation shown in Fig.~\ref{fig:heterogeneity}.
  }
  \label{fig:scars}
\end{figure}

The corresponding ventricular muscle fiber architecture was created for each individual ventricular geometry using a rule-based method developed by Quarteroni et al. \cite{Quarteroni2017}, see Fig.~\ref{fig:geometry_and_fibers}b).
Fiber and sheet directions were assigned to each particle or element as part of the initialization process of each simulation.
The algorithm ensures the rotation of the fiber and sheet directions through the ventricular wall by defining a rotation angle $\theta$ with
\begin{eqnarray}
    \theta &=& (\theta_{epi} - \theta_{endo}) \phi + \theta_{endo}
\end{eqnarray}
where the transmural position $\phi$ varies between 0 and 1, with $\phi = 0$ corresponding to the endocardial surface and $\phi = 1$ corresponding to the epicardial surface. 
This approach allows for a smooth and continuous change in fiber orientation throughout the myocardial wall, see Fig.~\ref{fig:geometry_and_fibers}b). 
We randomly chose a value for $\theta_{epi}$ from the range $-90^{\circ}$ to $-60^{\circ}$ and for $\theta_{endo} - \theta_{epi}$ from the range $80^{\circ}$ to $120^{\circ}$ from a uniform distribution for each simulation.

\begin{figure*}[htb]
  \centering
  \includegraphics[clip, trim=0.0cm 0.0cm 0.0cm 0.0cm, width=0.9\textwidth]{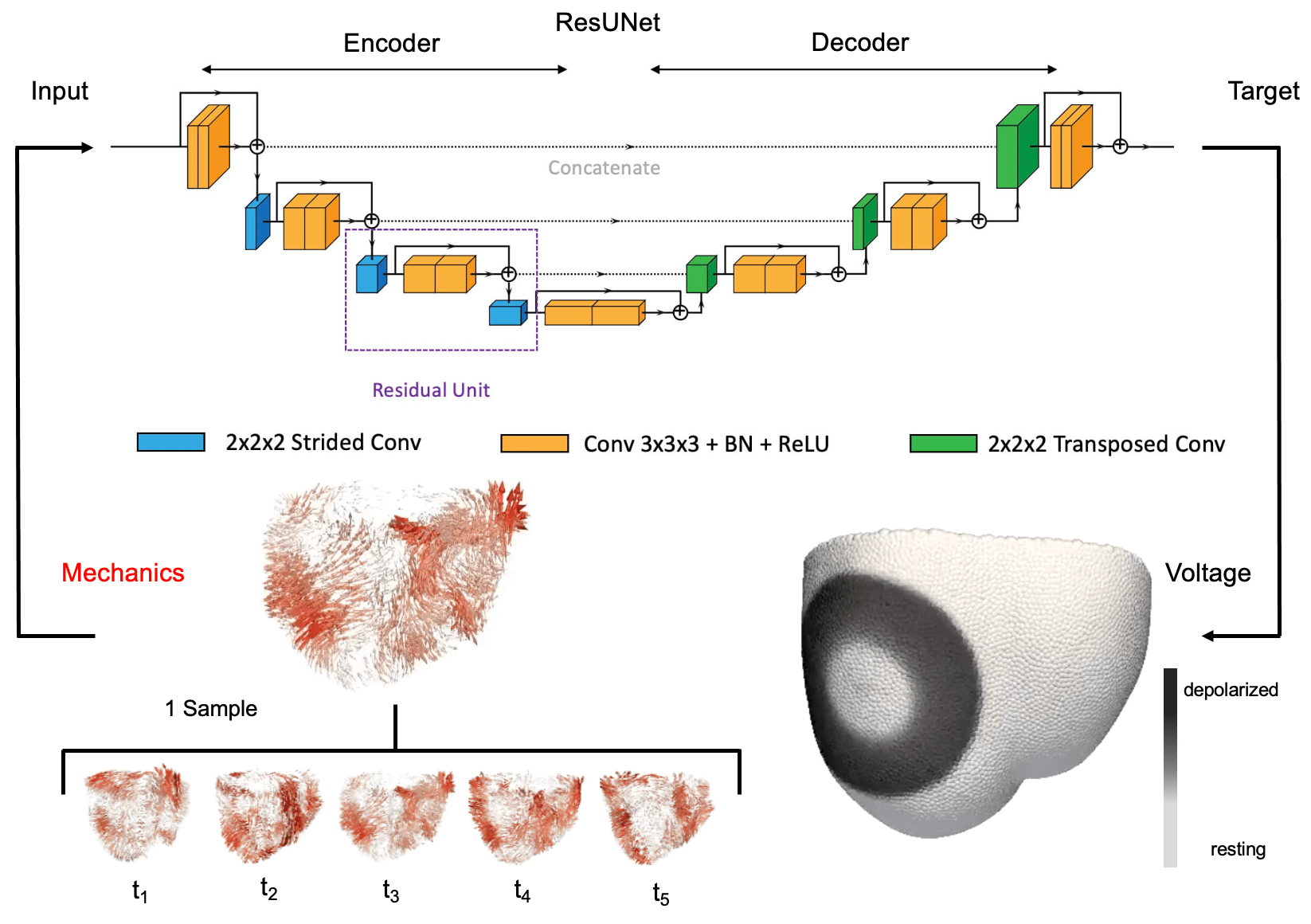}
  \caption{Neural network architecture and training samples for inverse mechano-electrical prediction of electrical arrhythmia circuits from ventricular deformation.
  The network architecture is a sparse fully convolutional ResUNet based on the U-Net architecture with encoding and decoding stage and skip connections as well as residual skip connections.
  The input to the network (left) is a short spatio-temporal sequence of displacement vector fields (here shown for 5 snapshots). 
  Each vector indicates the motion of a tissue particle in three-dimensional space. 
  The target or output of the network (right) during training or inference, respectively, is a three-dimensional scalar-valued field with normalized values for the transmembrane voltage for each tissue particle.}
  \label{fig:resunet}
\end{figure*}

\subsubsection{Spatial Heterogeneity Generation}
\label{sec:methods:heterogeneity}
Cardiac tissue frequently exhibits non-homogeneous dynamical properties that vary spatially throughout the heart.
To account for these variations and to generate a broader range of electrical dynamics, we employed three-dimensional Perlin noise \cite{Perlin1985,Perlin2002} in the SPH simulations with the goal to create smooth, spatially varying and unique noise patterns in each simulation, see also \cite{Jakes2019}. 
These patterns were then used to perturb the electrical dynamics, and indirectly the mechanical dynamics, simulating the inherent heterogeneity of the heart. 
We used Perlin noise to modulate the following two quantities in space: i) the diffusion process, ii) the dynamics of the slow/gating variables of the electrophysiological models.

To introduce spatial heterogeneity in the diffusion process, we implemented a spatially varying conductivity coefficient $c$, which scales the local diffusion rates, introducing spatial heterogeneity in the propagation of action potential waves. 
The $3\stimes3$ diffusion matrix $\textbf{D}_i$ for a particle $i$ is defined as:
\begin{align}
    \textbf{D}_i = c_i (D_{\text{iso}} \textbf{I} + D_{\text{aniso}} \textbf{f}_0 \otimes \textbf{f}_0)\text{,}
\end{align}
where $\textbf{I}$ is the identity matrix, $\textbf{f}_0$ is the local fiber direction, $\otimes$ is the outer product, and $D_{\text{iso}}$ and $D_{\text{aniso}}$ represent the isotropic and anisotropic components of the diffusion coefficient, respectively. 
The scalar value $c_i$ was determined at simulation initialization by the spatially varying Perlin noise:
\begin{align}
    c_i = 1 + \xi_h \xi_i\text{,}
\end{align}
where $\xi_h$ is a global scaling factor and $\xi_i = \xi_{\text{Perlin}}(\textbf{x}_{i,t=0})$ is the Perlin noise value for particle $i$ with initial position $\textbf{x}_{i,t=0}$. In our simulation, we vary the $D_{\text{iso}}$ and $D_{\text{aniso}}$ diffusion coefficients. For this we define:
\begin{align}
    D_{\text{ratio}} = \frac{D_{\text{iso}} - D_{\text{aniso}}}{D_{\text{iso}}}\text{.}
\end{align}
We randomly chose $D_{\text{ratio}}$ from the range 2.0 to 4.0 with a uniform distribution. For $D_{\text{iso}}$, we chose a value of $0.1\,\text{mm}^2/\text{ms}$ for reentrant simulations and randomly from a range $0.2$ to $0.4\,\text{mm}^2/\text{ms}$ for the focal simulations.

To alter the wave dynamics further, we introduced random spatial heterogeneity in the slow/gating variable(s), see also eq. (2), using Perlin noise:
\begin{align}
  \partial_t v \rightarrow (1 + \xi_h \xi_i) \partial_t v,
\end{align}
with $\xi_h$ and $\xi_i$ as above.
We chose a fixed block size or spatial frequency, comparable to the one shown in the center panel of Fig.~\ref{fig:heterogeneity}b), and modulated the strength of the noise ($\xi_h$) from $0.0$ to $0.2$.
Each realization of Perlin noise ($\xi_i$) used for the diffusion, the slow/gating variables as well was the scar generation described in the next section was different.

 \subsubsection{Scar Tissue Generation}
 \label{sec:methods:scars}
We modelled scar or infarcted tissue in the SPH simulations by procedurally generating areas of non-conducting and non-contracting tissue using Perlin noise as described in the previous section.
A threshold was defined to binarize the Perlin noise pattern, resulting in a spatial pattern consisting of ones and zeros, ones representing scar tissue.
As a result, scars were placed in random locations in the ventricles.
We varied the scar-to-healthy ratio or scar density from 0\%-20\%, see Fig.~\ref{fig:scars} and top right panel in Fig.~\ref{fig:parameters}.
A different realization of Perlin noise was used for the scar generation than for the spatial heterogeneity generation described in the previous section.

\subsection{Training Data Generation}
We generated two training datasets each including 40,000 training samples of pairs of electrical activation sequences and mechanical deformation during focal and reentrant arrhythmias using the data generated with the smoothed particle hydrodynamics (SPH) method, see section \ref{sec:methods:SPH}.
The data generated with the finite element method (FEM), see section \ref{sec:methods:FEM}, was excluded from the training data.
Each simulation has a unique set of parameters. 
One dataset excluded scar tissue, the other included scars.
For each of the 4 electrophysiological models (AP, FK, MS, BOCF), we produced 125 simulations for both focal and reentrant dynamics, in total $2 \cdot 4 \cdot 125 = 1,000$ simulations per dataset.
The focal and reentrant simulations were run for 2.6 s and 3.9s (simulation time), respectively. 
If the excitation of a reentrant simulation ceased completely, we restarted the simulation. 
To mimic various imaging speeds, we saved the simulation state every $\Delta t_{save}$ seconds (simulation time).
In each simulation, we randomly chose $\Delta t_{save}$ from the range 13\,ms to 32\,ms. 
This would correspond to imaging acquisition rates of 31\,Hz to 77\,Hz during 4D ultrasound imaging.
We extracted several unique training samples from one of the SPH simulations.
One training sample consists of 5 three-dimensional displacement vector fields describing the deformation of the ventricles and one corresponding three-dimensional set of points describing the cellular transmembrane voltage or three-dimensional electrical wave pattern.  
We used data augmentation to increase this initial training dataset size.
Data augmentation consists of rotating, scaling, cropping, subsampling the data (lowering the resolution), and adding (Gaussian) noise to the positions or displacement vectors.
The augmentations were equally applied to the vector-valued input data and the scalar-valued target data.
Additional simulations were performed for the generation of completely separate validation datasets.
The validations shown in Figs.~\ref{fig:accuracies} and \ref{fig:parameters} include 5,000 samples.

\subsection {Neural Network Architecture}
We developed a sparse fully convolutional encoder-decoder neural network to predict a scalar-valued pointcloud representing cellular transmembrane voltage from a sequence of vector-valued pointclouds representing a short spatio-temporal sequence of tissue motion, as shown in Fig.~\ref{fig:resunet}.
To improve the regularization and performance of the network, we included U-Net-like long skip connections \cite{Ronneberger2015} and ResNet-like short skip connections \cite{He2016}, also known as residual blocks.
To reduce memory and computational requirements, the neural network operates on sparse tensors instead of dense tensors utilizing the Minkowski Engine library \cite{Choy2019} implemented in PyTorch \cite{Paszke2019}.
The Minkowski Engine voxelizes the pointcloud input data into a sparse tensor and performs sparse convolutional operations.

In more detail, our neural network consists of an encoding part and a decoding part with a latent space in between.
The input to the neural network is a sparse tensor with $N$ channels generated by the Minkowski Engine using a quantization of 1.
The main building blocks of each part are residual blocks consisting of two 3D padded convolutional layers with a kernel size of $3 \times 3\times 3$, rectified linear unit (ReLU) activation \cite{Nair2010} and batch normalization \cite{Ioffe2015}.
The input of a residual block is passed through a convolutional layer with kernel size $1\times 1 \times 1$ and is added up element-wise to the output of the second 3D convolutional layer to form the residual connection. 
This short skip connection can improve performance and help to alleviate vanishing gradient problems \cite{He2016}. 
Additionally, long skip connections between encoding and decoding residual blocks are added by concatenating the output of the encoding block to the input of the decoding block. 
The encoding part consists of three residual blocks with 64, 128, and 256 filters for each convolutional layer. 
In the decoding part, we used three residual blocks with the same number of filters in reverse order, 256, 128, and 64. 
In the encoding part, we used a strided convolutional layer with kernel size 2×2×2, a stride of 2×2×2 after each block to downsample the input by a factor of two. 
Conversely, decoding residual blocks are processed by a transposed convolutional layer with the same parameters as the downsampling layers for upsampling. 
The downsampling and upsampling layers use the same number of filters as their respective residual block. 
The final layer of the decoding part is a convolutional layer with kernel size 1×1×1 with sigmoid activation and a single filter to map the output to a single scalar value. 
In this study, we used $N=5$ input channels as we intended the network to analyze 5 subsequent deformation states of the tissue. 
Accordingly, because the input is 5 three-dimensional vector-valued pointclouds, per point, the network receives $5 \cdot 3 = 15$ values as input ($x$-, $y$- and $z$-components) and maps these values to a single scalar.
We found empirically that 5 temporal snapshots provide sufficient information about the dynamics, see also \cite{Christoph2020,Lebert2021,Lebert2023}.
In total, our neural network has 27 million trainable parameters.

Training was performed on a NVIDIA RTX A6000 graphics processing unit in PyTorch \cite{Paszke2019} version 1.12.1 with mean squared error (MSE) as the loss function, the AdamW optimizer \cite{Loshchilov2019}, a learning rate of 0.001 and a weight decay of 0.01.
The training took about 60 hours with 40,000 samples for 64 epochs with a batch size of 32 samples.

%% file: sections/results.tex
We found that it is possible to predict focal and reentrant electrical wave patterns in simulated bi-ventricular heart geometries from the deformation that occurs in response to the electrical activation. 
The predictions can be performed in unseen, arbitrarily-shaped ventricular geometries with complex muscle fiber architecture, even with significant tissue heterogeneity and the partial presence of non-conducting and non-contracting scar tissue.

\begin{figure}[htb]
  \centering
  \includegraphics[clip, width=0.49\textwidth]{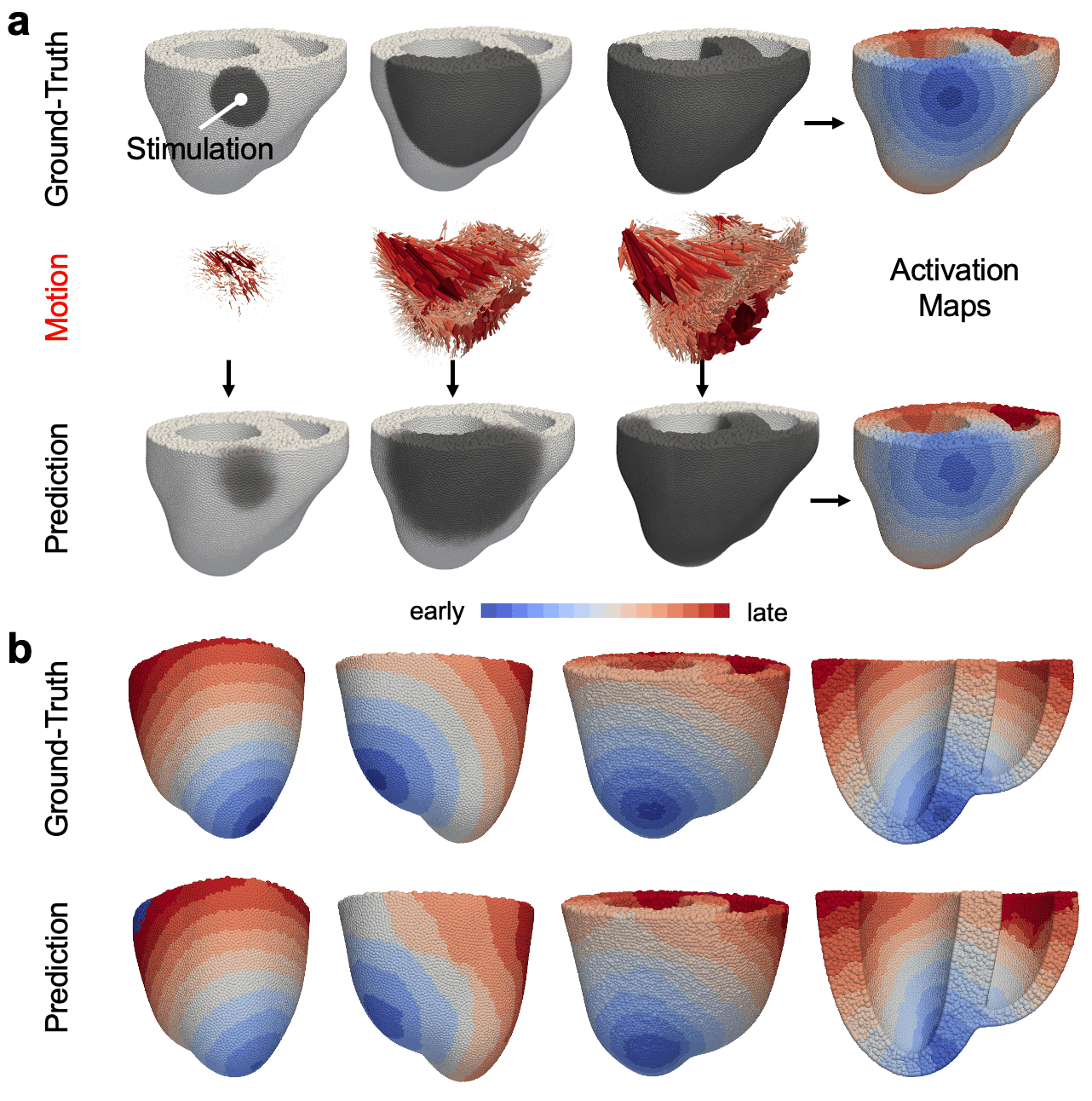}
  \caption{Electrical activation maps computed from ventricular motion using deep learning, see also Supplementary Video 4.
  a) Top: Ground-truth focal electrical wave pattern (dark: depolarized tissue, gray: resting tissue) with corresponding electrical activation map (blue: early activation, red: late activation).
  Center: Motion of the ventricles in response to electrical activation.
  Bottom: Electrical wave pattern predicted from ventricular motion and corresponding electrical activation map computed from the sequence of predictions.
  b) Different examples of electrical activation maps computed from ground-truth (top) and predicted (bottom) dynamics.
  The particular ventricle geometries and activation patterns were not seen by the neural network during training.
  }
  \label{fig:predictions_focal_activation}
\end{figure}

\subsection{Focal Waves}
\label{sec:results_focal}
\begin{figure*}[htb]
  \centering
  \includegraphics[clip, trim=0.0cm 0.0cm 0.0cm 0.0cm, width=0.92\textwidth]{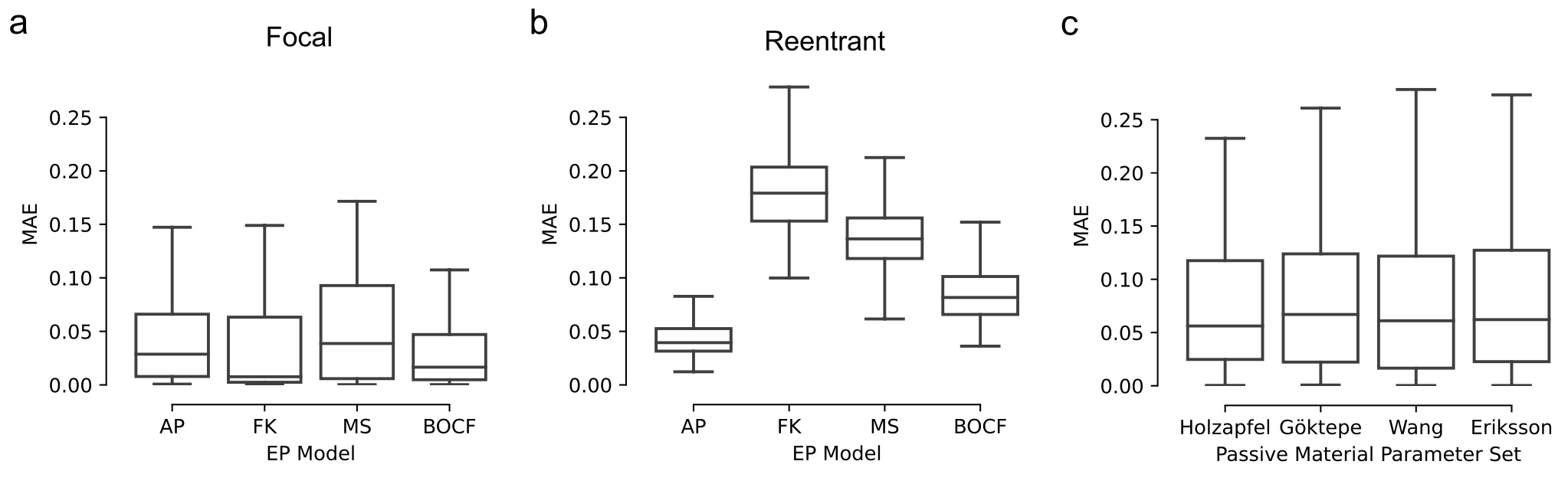}
  \caption{Prediction accuracies of deep learning-based inverse reconstruction of electrical waves from ventricular deformation of the heart on synthetic simulated data generated with different electrical and mechanical models.
 Prediction accuracies for a) focal and b) reentrant waves generated with either of the 4 electrophysiological models (AP: Aliev-Panfilov \cite{AlievPanfilov1996}, FK: Fenton-Karma \cite{FentonKarma1998}, MS: Mitchell-Schaeffer  \cite{MitchellSchaeffer2003}, BOCF: Bueno-Orovio-Cherry-Fenton \cite{BuenoOrovioCherryFenton2008}).
 The training data contains all 16 combinations of electrical and mechanical models, varying parameters and scars.
 While the prediction error (MAE: mean absolute error) stays small in the order of 5\% for all models with focal waves, reentrant waves produce higher prediction errors, particularly the ones with more complicated scroll wave dynamics.
 c) Prediction accuracies on a mix of focal and reentrant data generated with 4 different mechanical material properties (\textcite{Holzapfel2009}, \textcite{Goktepe2010b}, \textcite{Wang2012}, \textcite{Eriksson2013}), see section \ref{sec:methods:SPH}.
 The data suggests that the neural network performs equally well across different electrical and mechanical models, unless the dynamics are more complex (as with the Fenton-Karma or Mitchell-Schaeffer models).
 All errors are the mean absolute error (MEA) computing the difference between ground-truth and predicted electrical wave patterns over all validation samples.
  }
  \label{fig:accuracies}
\end{figure*}

Focal waves can cause premature ventricular contractions (PVCs) and may underlie slower hemodynamically stable monomorphic ventricular tachycardia (VT).
PVCs commonly originate from the papillary muscles, but can also originate from other locations in the ventricles.
Fig.~\ref{fig:predictions_focal} shows that our deep learning algorithm can identify focal electrical wave patterns from the ventricular deformation that occurs in response to these focal wave patterns.
The predictions show focal electrical wave patterns that propagate outwards and away from their focal origin, see also Supplementary Videos 2 and 3.
The algorithm produces a sequence of predictions which is temporally smooth and consistent over time, even though each prediction was performed individually without knowledge about the electrical wave patterns in previous time steps.
Panels a) and b) in Fig.~\ref{fig:predictions_focal} show two representative examples with shorter and longer wavelengths, and we found that the predictions perform equally well across all wave regimes, electrical and mechanical models and model parameters, see also Fig.~\ref{fig:accuracies}a,c), and the focal waves can originate from any random location in the ventricles.
The predicted electrical wave patterns are three-dimensional and transmural.
Overall, the prediction accuracies with focal waves are in the order of 90-95\%, see Figs.~\ref{fig:accuracies}a) and \ref{fig:parameters}, across all combinations of the electrical and mechanical models and parameters.

\begin{figure*}[!ht]
  \centering
  \includegraphics[clip, trim=0.0cm 0.0cm 0.0cm 0.0cm, width=0.98\textwidth]{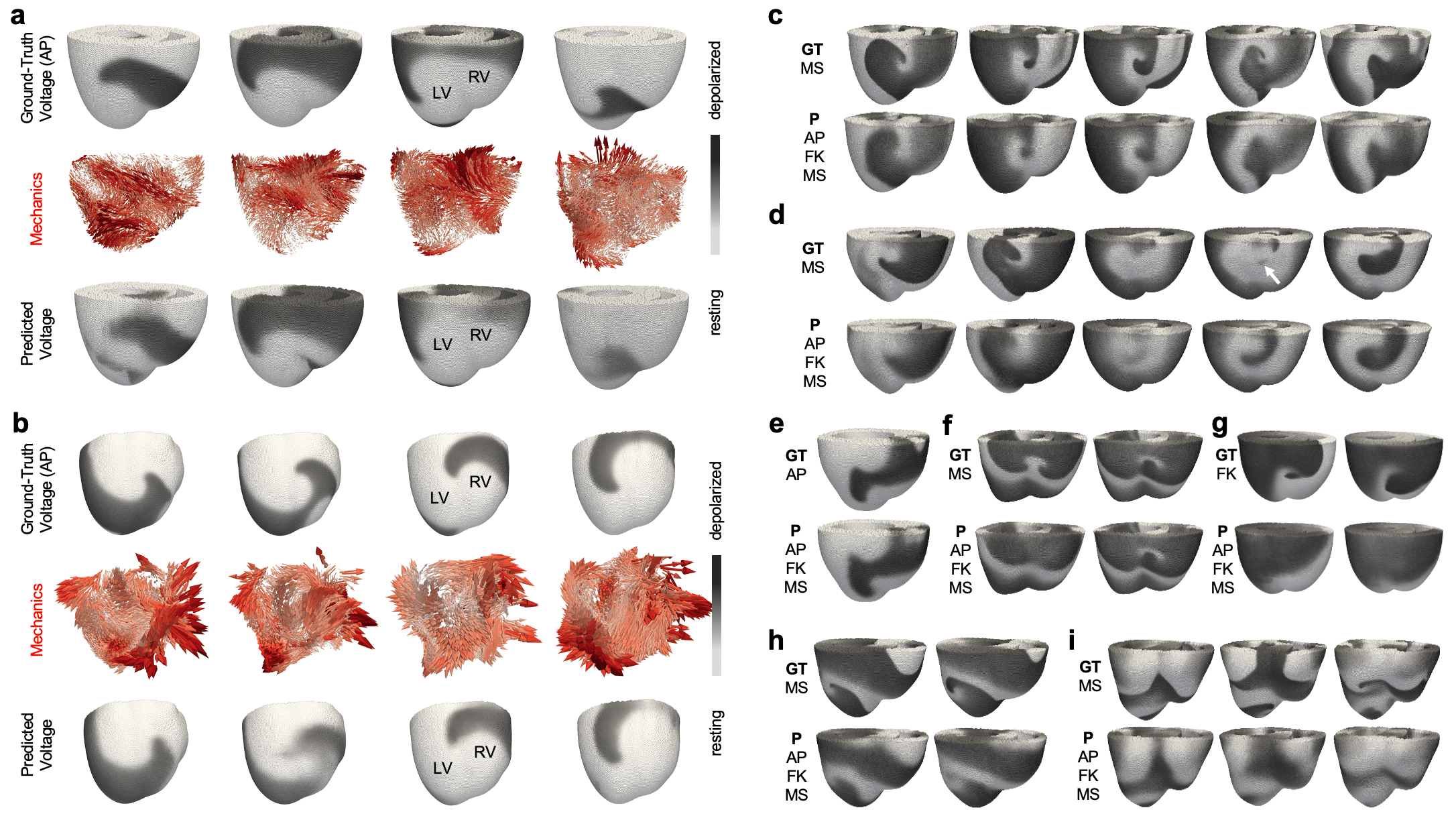}
  \caption{Deep learning-based predictions of reentrant electrical (action potential) waves from ventricular motion, see also Supplementary Videos 5-18. 
  The deep learning model was trained on reentry data generated with the Aliev-Panfilov (AP) \cite{AlievPanfilov1996}, the Mitchell-Schaeffer (MS) \cite{MitchellSchaeffer2003} and the Fenton-Karma (FK) \cite{FentonKarma1998} electrophysiology models without scars but with significant spatial heterogeneity in electrical and mechanical model parameters. 
  a,b) Counter clock-wise rotating electrical scroll waves, corresponding mechanical deformation of the ventricles and predictions. 
  c) Clock-wise rotating scroll wave (GT: ground-truth) and prediction (P) generated with MS model.
  d) Epicardial breakthrough generated with MS model.
  e,f) Double rotors or figure-of-eight patterns.
  g) Poor prediction outcome with FK model.
  h) Predictions are low-pass filtered versions of GT wave pattern.
  i) Predictions of fully turbulent scroll wave dynamics.
  The (GT) data on which the predictions were performed was not seen by the neural network during training (neither the particular bi-ventricular geometry nor the reentrant waves). 
  The electrical and mechanical tissue properties were different in each simulation.}
  \label{fig:predictions_reentrant}
\end{figure*}

Using the sequence of predictions, it is possible to locate the origin of focal electrical waves within the ventricles.
Fig.~\ref{fig:predictions_focal_activation} shows electrical activation maps, which were 
computed from the sequence of predicted electrical wave patterns, and therefore computed from the ventricles motion.
Panel a) shows a sequence of a ground-truth and predicted focal electrical wave patterns and the two corresponding electrical activation maps, which are visually nearly identical, see also Supplementary Video 4.
Panel b) shows other representative examples of ground-truth and predicted activation maps, one of which shows an intramural wave source approximately at midwall close to the apex (right).

\begin{figure*}[htb]
  \centering
  \includegraphics[clip, trim=0.0cm 0.0cm 0.0cm 0.0cm, width=0.96\textwidth]{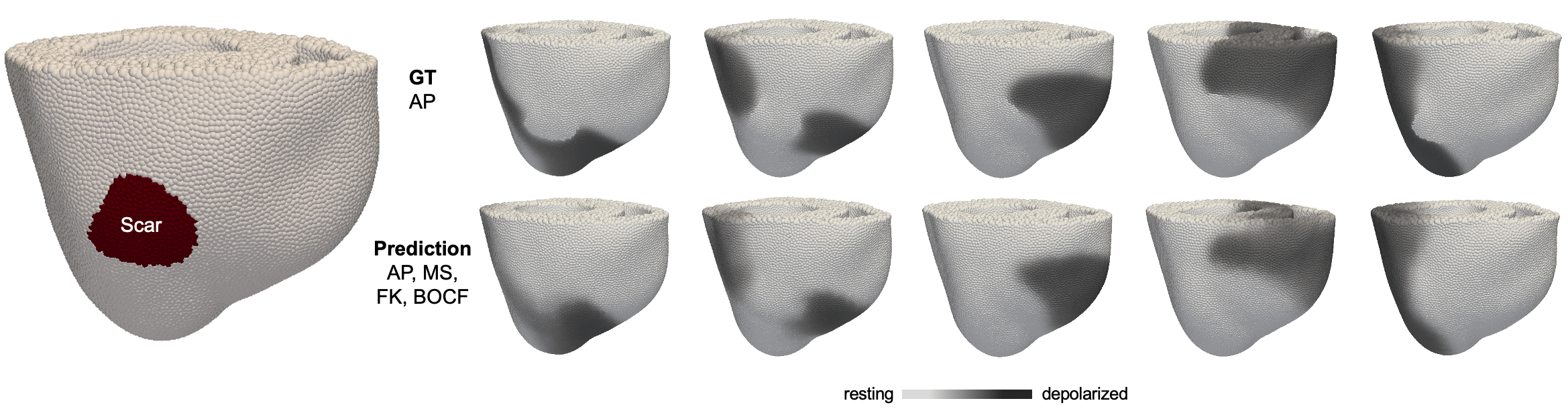}
  \caption{Deep learning-based prediction of electrical (action potential) waves from ventricular deformation in the presence of scar tissue, see also Supplementary Videos 19 and 20.
  A random number of scars were placed in random locations in the ventricles (left, deep red). 
representative example of a reentrant electrical wave generated with the Aliev-Panfilov model rotating around scar in the left ventricle (top: ground-truth / GT, bottom: prediction). 
The prediction does not exhibit electrically activated tissue in the scar region.
The network was trained on 4 electrophysiological models (AP, MS, FK, BOCF) with focal and reentrant waves and scars with a healthy-to-scar tissue ratio ranging from 0-20\%, see also Fig.~\ref{fig:scars}.
  The data on which the predictions were performed was not seen by the neural network during training.}
  \label{fig:predictions_scars}
\end{figure*}

The examples shown in Figs.~\ref{fig:predictions_focal} and \ref{fig:predictions_focal_activation} were not part of the training data and were therefore not seen by the neural network during training.
In particular, neither the bi-ventricular geometries nor the particular focal waves with their origin, wavelength and conduction velocity were part of the training data.
Focal waves can also be predicted with an unknown electromechanical computer model, see Fig.~\ref{fig:fem}a).
While the examples in Figs.~\ref{fig:predictions_focal} and \ref{fig:predictions_focal_activation} do not include any scar tissue, we found that predictions of focal waves are also possible in the presence of scars with slightly lower accuracies, see section \ref{sec:results:scars} and Figs.~\ref{fig:predictions_scars}a) and \ref{fig:accuracies}a) as well as the top right panel in Fig.~\ref{fig:parameters}.

\subsection{Reentrant Waves}
\label{sec:results_reentrant}
Reentrant waves underlie faster and hemodynamically unstable monomorphic or polymorphic ventricular tachycardia (pVT), Torsade de Pointes (TdP) or ventricular fibrillation (VF). 
Fig.~\ref{fig:predictions_reentrant} shows that it is possible to reconstruct three-dimensional reentrant electrical wave patterns from the ventricular motion that these patterns caused using our deep learning approach.
Panels a) and b) show reentrant electrical scroll waves produced with the Aliev-Panfilov model (AP) \cite{AlievPanfilov1996}, their corresponding mechanical ventricular deformations and predictions.
The predictions (P) exhibit minor artifacts, but are very similar to the original ground-truth (GT) wave pattern. 
The sequence of predictions is temporally smooth and consistent over time, see also Supplementary Videos 5-18, even though the predictions were performed independently from each other without knowledge about the electrical waves in previous time steps.
The waves in a,b) are idealized for illustration purposes and are relatively `well-behaved'.
Panel c) shows a more chaotic and intricate clock-wise rotating reentrant wave produced with the Mitchell-Schaeffer model (MS) \cite{MitchellSchaeffer2003}.
The wave meanders very strongly and evolves quickly into a completely different wave pattern after a few rotations.
The dynamics are more comparable to pVT, TdP or VF. 
Overall, the deep learning model reconstructs a blurred version of the spiral-like wave pattern that is consistent with the global wave pattern throughout the volume of both ventricles, but lacks to reproduce finer details of the wave pattern.
Panel d) shows another clock-wise rotating wave generated with the MS model, which breaks through the epicardial surface (white arrow). 
The prediction is able to recapitulate this breakthrough, indicating that the approach is able to resolve transmurality, but struggles with finer details of the wave pattern.
Panels e) and f) show successful predictions of double rotors or figure-of-eight patterns with the AP and MS models, respectively. 
Here the model succeeded with reproducing some of the finer details of the wave pattern.
Panel g) shows a large clock-wise rotating reentrant wave generated with the Fenton-Karma model (FK) \cite{FentonKarma1998}.
Overall, the deep learning model performed poorly with reentrant waves generated with the FK model, presumably because it exhibited the most diverse dynamics among the 4 electrophysiology models. 
The FK model produced both larger macro-reentrant waves as well as very fine-scaled VF-like waves, and the deep learning model might require more training data.
Panel h) shows another example with the MS model in which the predictions appear like low-pass filtered versions of the ground-truth wave pattern.
Panel i) shows predictions of 'fully turbulent' scroll wave dynamics. Again, the predictions capture the global wave patterns but smear out finer details.
The data suggests that with further improvements our approach could even be used to reconstruct three-dimensional action potential wave patterns underlying VF.

In summary, it is possible to reconstruct even complicated reentrant wave patterns from ventricular motion.
The prediction accuracy is in the order of 80-95\%, see Fig.~\ref{fig:accuracies}, depending on whether the wave dynamics are simpler or more complex VF-like and depending on which electrophysiological model was used.
The deep learning model used in Fig.~\ref{fig:predictions_reentrant} was trained with reentrant data generated with the AP, FK and MS models without scars, but with significant spatial heterogeneity in the electrical and mechanical parameters, see also Fig.~\ref{fig:predictions_scars}.
Likewise, all predictions were performed without scars, but with heterogeneity.
Fig.~\ref{fig:predictions_scars} shows predictions in the presence of scar tissue, see also section \ref{sec:results:scars}.

All predictions were performed on unseen data that was not part of the training data.
However, the analyzed data was generated with electrical and mechanical models also used to generate the training data (AP, MS, FK and their combinations with one of the mechanical models).
Consequently, the data to which the deep learning model was applied stems from within the training distribution, and the dynamics were in principle already known to the deep learning model from the training data.
Fig.~\ref{fig:fem} shows predictions on data that does not come from the same training data distribution, see also section \ref{sec:results:generalization}.

\subsection{Heterogeneity and Scar Tissue}
\label{sec:results:scars}
\begin{figure*}[htb]
  \centering
  \includegraphics[clip, trim=0.0cm 0.0cm 0.0cm 0.0cm, width=0.94\textwidth]{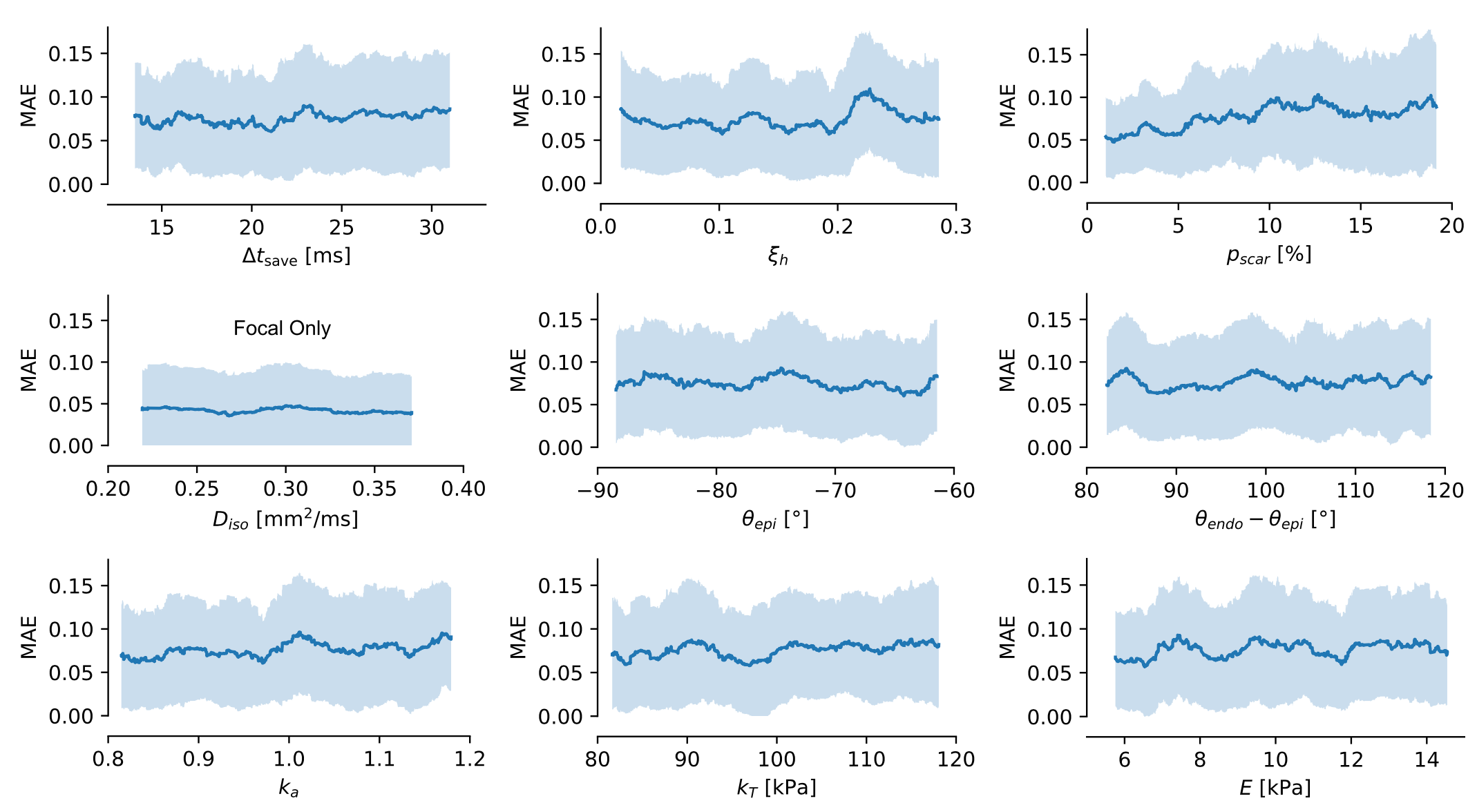}
  \caption{The prediction accuracy of the neural network stays constant over the range of each of the varied parameters. On average, the error stays smaller than 90\% except with scars.
 All errors are the mean absolute error (MEA) computing the difference between ground-truth and predicted electrical wave patterns over all validation samples and for a mix of focal and reentrant waves with scars.
Top left: $\Delta t_{save}$ defines the sampling rate used to generate a training sample and relates to imaging speed.
Top center: $\xi_h$ defines the strength of the spatial heterogeneity parameter.
Top right: $p_scar$ defines the ratio of scar-o-healthy tissue. The prediction error increases slightly from 5\% to 10-15\% with increasing scar size. 
Center left: $D_{iso}$ is the diffusion constant which generally influences wavelength and speed.
Center: $\theta_{epi}$ defines the fiber angle on the epicardium to initialize the rule-based generation of the fiber alignment in the ventricles.
Center right: $\theta_{endo}-\theta_{epi}$ defines the total fiber angle change through the wall.
Bottom left: $k_a$ modulates the passice material properties, see Table \ref{tab:mechanics_parameters} and section \ref{sec:methods:SPH}.
Bottom center: $k_T$ is a parameter influencing the contractile strength, see eq. (6).
Bottom right: $E$ is the Young's modulus.
  }
  \label{fig:parameters}
\end{figure*}

Tissue heterogeneity or scar tissue does not pose a limitation to the inverse mechano-electrical neural network.
Fig.~\ref{fig:predictions_scars} demonstrates that electrical waves can be predicted from tissue motion even in the presence of scars and with spatial heterogeneity.
Scars were simulated as non-conducting and non-contracting tissue and were placed in random locations in the ventricles (red). 
The representative example in Fig.~\ref{fig:predictions_scars} shows a reentrant wave generated with the Aliev-Panfilov (AP) model rotating in counter clock-wise direction around a scar. 
Supplementary Videos 19 and 20 show additional examples of predictions with scars, and we obtained similar results with focal waves with the Mitchell-Schaeffer (MS) and Bueno-Orovio-Cherry-Fenton (BOCF) models, respectively.
The predictions exhibit rarely electrical excitation inside scarred areas.
However, they also exhibit blurring of the electrical wave patterns in and around the scar.
Overall, the prediction accuracies with scars are in the order of 90\% with scars. 
Panel a) in Fig.~\ref{fig:accuracies} shows prediction accuracies in the order of 95\% with focal waves and scars with all 4 electrophysiology models (AP, FK, MS, BOCF, evenly split training mix).
Accordingly, panel b) shows prediction accuracies with reentrant waves in tissues with scars.
The prediction accuracy varies from about 95\% with the AP model to 80\% with the Fenton-Karma model (FK).
There is no significant difference with different mechanical models (material properties) on the prediction accuracy, as shown in panel c), when evaluating on a mix of both focal and reentrant waves.
Prediction accuracies decrease with increasing scar size or scar content (scar-to-healthy tissue ratio), see top right panel in Fig.~\ref{fig:parameters}, but become only marginally worse with scars overall.
Spatial heterogeneity of tissue parameters was present in all simulations throughout this study, as shown in Fig.~\ref{fig:heterogeneity}, independently from scars.
Spatial heterogeneity in tissue parameters was part of all training data and did not have an effect onto the prediction accuracy, see top center panel in Fig.~\ref{fig:parameters}.
All other parameters in Fig.~\ref{fig:parameters} did not influence the prediction accuracy (MEA: mean absolute difference).
The data on which the predictions were performed was not seen by the neural network during training (neither the particular bi-ventricular geometry nor the scars nor the wave patterns).
Both training data and data on which predictions were performed in Figs.~\ref{fig:predictions_scars}, \ref{fig:accuracies} and \ref{fig:parameters} contained varying degrees of scar sizes and spatial heterogeneity. 
In summary, our deep learning-based algorithm can reliably predict electrical wave patterns from ventricular motion even in the presence of significant heterogeneity and non-conducting and non-contracting tissue.

\subsection{Generalization}
\label{sec:results:generalization}
\begin{figure}[ht]
  \centering
  \includegraphics[clip, trim=0.0cm 0.0cm 0.0cm 0.0cm, width=0.49\textwidth]{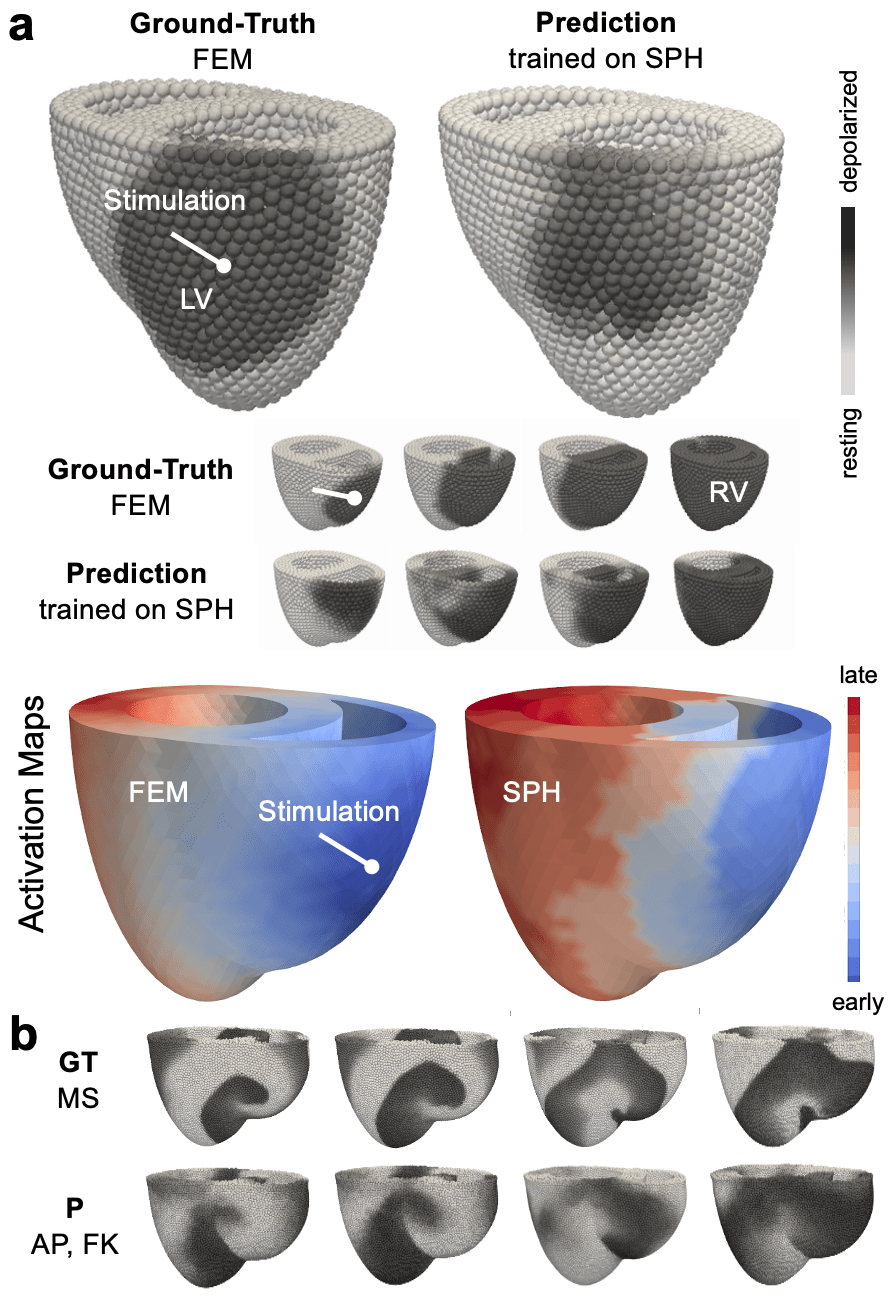}
  \caption{Generalization of neural network to out-of-distribution data.
 a) Deep learning model trained on SPH data applied to FEM data. 
 The essence of the focal wave that was simulated using FEM can be recovered even though the network only knows focal dynamics from the SPH method. 
b) Reentrant electrical wave (GT: ground-truth) simulated with the SPH method and Mitchell-Schaeffer model (MS) reconstructed using a deep learning model trained on data generated with the SPH method and the Aliev-Panfilov (AP) and Fenton-Karma (FK) models. 
  }
  \label{fig:fem}
\end{figure}

One of the most critical questions regarding deep learning models is whether they generalize to distinctly different data, or whether they produce reliable predictions when applied to data that is well outside of the training data distribution, see Fig. \ref{fig:fem}.
We refer to this as applying the network to `out-of-distribution' data.
In the clinic, a deep learning model should perform sufficiently well on data that was obtained in a different patient population not included in the training data and is therefore unknown to the model.
To test this behaviour, we applied our deep learning model to data that was distinctly different from the training data.

Most importantly, we trained our deep learning model exclusively on data generated with SPH and subsequently applied it to data generated with FEM, see Fig.~\ref{fig:fem}a).
The physics of the FEM model was not known during the development of the training dataset, and therefore the SPH simulations could not be calibrated to match the FEM simulations. 
As a result, the data produced with the two simulations was quite different. 
For example, the mechanical boundary conditions of the SPH and FEM implementations were different in that the base of the ventricles in the SPH simulations was attached to an elastic medium that was fixed in space, whereas in the FEM simulations it was fixed in vertical direction but could rotate in the horizontal plane.
In Fig.~\ref{fig:fem}a), it is shown that the neural network was able to predict the evolution of focal electrical waves from FEM data, even though it was trained solely on SPH data and therefore unfamiliar with the electromechanical dynamics of the FEM model. 
Despite the differences in the SPH and FEM modelling, the neural network was able to predict focal waves originating in the left ventricle (LV), right ventricle (RV) and septum, even though the mechanical behavior and boundary conditions were different and the FEM simulations were performed at lower resolutions (we tested N=6 examples: LV/RV/septum low resolution, LV/RV/septum very low resolution).
The predictions exhibit stronger artifacts and are less accurate than with SPH data, but capture the essence of the electrical activation patterns, which is reflected by the similarity between the two activation maps.
We expect that the prediction quality will improve if the SPH simulations are further refined to match the FEM simulations or vice versa (e.g. matching mechanical boundary conditions).
In our case, however, the mismatch between the two methodologies serves the purpose to demonstrate the generalization capabilies of the network.
The model was trained on both focal and reentry waves as well as with scars.

We also trained our deep learning model on data generated solely with the SPH method and with a subset of the 4 electrophysiological models (leaving one of the models out of the training data), and then applied it to data generated with a model that was not used during training.
Fig.~\ref{fig:fem}b) shows such an `out-of-distribution' prediction with reentrant electrical waves, where the (GT: ground-truth) data, which the deep learning model analyzed to reconstruct the electrical wave, was generated with the Mitchell-Schaeffer (MS) model, and the deep learning model was trained with the AP and FK models.
Supplementary Video 21 shows predictions of reentrant scroll wave dynamics, the deep learning model analyzing data generated with the BOCF model (ground-truth), while trained on just the AP, MS and FK models.
Even though the predicted electrical waves are blurry, the overall shape and evolution of the predicted pattern matches the ground-truth and it is possible to identify a reentrant scroll wave pattern from the predictions.
Together, Fig.~\ref{fig:fem}b) and the video demonstrate that predictions can be performed, even if the deep learning model has never seen the particular electrical dynamics before.

Lastly, we tested our deep learning model on geometries which are completely different from the bi-ventricular shape.
Supplementary Video 22 shows predictions of electrical waves propagating through a teapot (the ``Utah teapot'') that deforms accordingly (we refer to this type of arrhythmia as ``VTea''). 
While the predictions do not perform as well as with the bi-ventricular geometries, presumably because the teapot has finer geometric details, they do produce wave patterns moving through the walls of the teapot.

Overall, the findings demonstrate that the neural network can perform predictions with mechanical data that lies well outside of the distribution of the training data, with arbitrary geometries (given the limitations regarding spatial resolution), and independently from the particular methodology used to create the data.
The findings suggest that the neural network has the capability to generalize, but also reveal limitations.

%% file: sections/discussion.tex
In this numerical study, we showed that deep convolutional neural networks are able to reconstruct three-dimensional electrical wave patterns from ventricular deformation.
After training our network with a diverse electromechanical dataset of simulated ventricular arrhythmias, it is able to devise a unique, generalizable mapping between mechanical deformation and electrical wave patterns in any ventricular geometry, despite the many factors potentially influencing this relationship: the infinitely many possible electrical wave patterns, the arbitrarily-shaped ventricular geometries, the long-range elastic forces acting throughout the tissue causing non-local interactions between tissue segments (tissue shape affecting mechanics and vice versa), the complex fiber architecture, the spatial heterogeneity in electrophysiological and mechanical parameters, the non-contracting heterogeneities in the tissue, etc..
Given the variety of electromechanical tissue dynamics and heterogeneity that we simulated, it is remarkable that the neural network was able to produce a universal mapping that translates any given mechanical spatio-temporal pattern into a corresponding electrical spatio-temporal pattern and can identify an underlying electrical arrhythmia circuit from any three-dimensional deformation pattern of the ventricles.
Our test on completely unknown data generated with a completely different methodology (FEM), see section \ref{sec:results:generalization}, suggests that the network has learned the essential features of the dynamics, generalizes to some extent, and could potentially be applied to any other data as well, maybe even experimental data.
At the same time, the network has no explicit knowledge about the underlying physics and parameters.
For instance, the fiber orientation did not need to be known to the network.
This could be a decisive advantage over physics-based approaches \cite{Otani2010, Lebert2019, Beam2020} when applying our approach to experimental data, for which such parameters are often unknown.

Our central approach was to use efficient phenomenological modelling to be able to generate an extensive and diverse electromechanical training dataset in hopes that the distribution of training samples would be so broad that it spans across data that the network would later encounter during application `in the wild'.
On the one hand, this approach was a necessity, because more realistic computer models of the contracting heart during arrhythmias are prohibitively expensive (computationally, at least as of now).
On the other hand, we intentionally used phenomenological modelling, because we believe that a broad training data distribution with exaggerated and idealized tissue shapes and dynamics will strengthen the network's robustness, promote generalization, and prevent overfitting.
While we desire to include training samples obtained with more realistic patient-specific simulations in the future, we believe that the idealized dynamics can already capture the most essential components of the relationship between electrics and mechanics. %
We expect that more detailed computer simulations would not alter the findings and implications of our study, but would help to translate our approach into the clinic.
Phenomenological simulations could help to make the training data generation computationally tractable and could provide a foundation for further fine-tuning. 

One of the main challenges in training our deep learning model is creating training data that contains an ensemble of dynamics, within which the dynamics are evenly distributed with as little bias as possible. 
How does one design such a training dataset?
On the one hand, there are the technical aspects of creating a uniform distribution of training samples in a hyper-dimensional parameter space. 
We varied 10-15 simulation parameters, often sampling each single parameter from a uniform distribution, which could lead to a heavily biased distribution of the dynamics, especially when varying model parameters.
The current literature does not address these issues.
On the other hand, there is the question about which arrhythmia dynamics to include in the training.
In our study, we saw that Fenton-Karma dynamics are difficult to reconstruct, presumably because there were not enough examples in the training dataset (we used an even split between models). 
Should certain more complex arrhythmias be favored over simpler ones?

Training will ultimately and ideally be performed with imaging data.
Experimental data of ventricular motion is available in the same form as shown in Figs.~\ref{fig:electromechanical_rotor}b) and \ref{fig:inverse-problem}: as a time-varying sequence of three-dimensional displacement vector fields that can be obtained when tracking motion in a sequence of three-dimensional images.
These images could be obtained with ultrasound as in \cite{Christoph2018} or magnetic resonance imaging, for example.
We mimicked different imaging speeds by providing training samples which cover either a shorter or a longer fraction of the cyclelength to the network.
Using this training strategy, the network subsequently learned to perform predictions independently from the imaging speed, which could become very relevant in practice.
The tissue displacements in imaging data can be computed either from one frame to the next or for all frames with respect to one of the frames. 
We found in \cite{Christoph2020} that both frames of reference perform equally well, as long as either approach is used consistently during training and prediction.
Noteworthy, the network does not need to know an undeformed relaxed tissue configuration, which means that any arrhythmic episode can be imaged.
Whether our network will be able to analyze imaging data needs to be determined in future work.

One interesting question is if it is possible to train only on simulations and directly apply the network to clinical data. 
This will likely depend on how far off the simulated data will be from the imaging data.
In a different study, we found that a neural network that was trained solely on simulated data can detect core regions or phase singularities of reentrant waves in optical mapping data \cite{Lebert2021}.
The mechanical data in our current training dataset contains noise and other augmentations, but imaging data may have more convoluted features that are not yet present in our training data.
Furthermore, simulated training data will require more physiological detail.
Currently, our modelling does not include the atria and does not account for many of the other physiological processes in the human body, such as blood flow.

In principle, our deep learning approach is not limited to predicting electrical waves from motion in the ventricles, but could also be applied in the atria and could furthermore be used to predict calcium waves, activation maps, locations of active stress, scars or other predictors for arrhythmias, if the network was trained accordingly, see also \cite{Christoph2020}.
For instance, training data for deep learning-based scar predictions could stem from either late gadolinium-enhanced magnetic resonance imaging, from simulations, or from physics-based approaches \cite{Otani2010,Kovacheva2021}.
Whether our approach will be feasible in the atria is difficult to anticipate, because their arrhythmia dynamics and anatomical, electrophysiological and mechanical properties are so different from the ventricles.
Because tissue heterogeneity does not appear to pose a limitation, our approach could eventually also be applied in heart failure patients \cite{Maffessanti2020}.
Lastly, a fundamental limitations for our approach could occur during the degeneration of the coupling between electrics and mechanics. 
In this study, we made a critical simplification in that we assumed that calcium waves always follow electrical waves.
This is not always the case, especially during atrial or ventricular fibrillation, during which the coupling between voltage and calcium can become bi-directional or otherwise degenerated \cite{Omichi2003,Warren2007,Weiss2010,Voigt2012,Greiser2010,Greiser2014}.
We aim to study the impact of such electromechanical dissociation phenomena in future work.

%% file: sections/conclusions.tex
We provided a numerical proof-of-principle that electrical arrhythmia circuits in the heart can be computed from ventricular deformation mechanics using deep learning.
We demonstrated that deep neural networks can learn to associate the deformation of arbitrarily-shaped ventricular tissues with the corresponding three-dimensional morphology of an electrical wave pattern that caused the deformation, regardless of whether it is a simpler focal or more complicated reentrant rhythm, even in the presence of heterogeneity and scar tissue.
The results suggest that a similar artificial intelligence-based approach could be used in combination with high-speed imaging, such as ultrasound or magnetic resonance imaging, to visualize action potential waves, activation maps and other electrical predictors non-invasively in patients with ventricular arrhythmias.
These findings require confirmation in imaging experiments and in more detailed, patient-specific simulations.